\documentclass[reprint,superscriptaddress,amsmath,amssymb,aps,nofootinbib,floatfix]{revtex4-1}
\usepackage{color,units}
\usepackage[colorlinks=true]{hyperref}
\usepackage{multirow}
\usepackage{graphicx}
\usepackage{lipsum}
\usepackage{dcolumn}
\usepackage{bm}
\usepackage{CJKutf8}
\usepackage[normalem]{ulem}

\begin{document}

\preprint{APS/123-QED}

\title{Inferring the population properties of binary neutron stars\\ with gravitational-wave measurements of spin}

\author{Xingjiang Zhu}
\email{xingjiang.zhu@monash.edu}
\affiliation{School of Physics and Astronomy, Monash University, Clayton, Vic 3800, Australia}
\affiliation{OzGrav: Australian Research Council Centre of Excellence for Gravitational Wave Discovery}
\author{Eric Thrane}
\affiliation{School of Physics and Astronomy, Monash University, Clayton, Vic 3800, Australia}
\affiliation{OzGrav: Australian Research Council Centre of Excellence for Gravitational Wave Discovery}

\author{Stefan Os{\l}owski}
\affiliation{Centre for Astrophysics and Supercomputing, Swinburne University of Technology, Hawthorn Vic 3122, Australia}
\affiliation{OzGrav: Australian Research Council Centre of Excellence for Gravitational Wave Discovery}

\author{Yuri Levin}
\affiliation{Physics Department and Columbia Astrophysics Laboratory, Columbia University, New York, NY 10027}
\affiliation{Center for Computational Astrophysics, Flatiron Institute, New York, NY 10010}
\affiliation{School of Physics and Astronomy, Monash University, Clayton, Vic 3800, Australia}

\author{Paul D. Lasky}
\affiliation{School of Physics and Astronomy, Monash University, Clayton, Vic 3800, Australia}
\affiliation{OzGrav: Australian Research Council Centre of Excellence for Gravitational Wave Discovery}

\date{\today}

\begin{abstract}
The recent LIGO-Virgo detection of gravitational waves from a binary neutron star inspiral event GW170817 and the discovery of its accompanying electromagnetic signals mark a new era for multimessenger astronomy.
In the coming years, advanced gravitational-wave detectors are likely to detect tens to hundreds of similar events.
Neutron stars in binaries can possess significant spin, which is imprinted on the gravitational waveform via the effective spin parameter $\chi_\text{eff}$.
We explore the astrophysical inferences made possible by gravitational-wave measurements of $\chi_\text{eff}$.
First, using a fiducial model informed by radio observations, we estimate that $\approx15-30\%$ of binary neutron stars should have spins measurable at $\gtrsim 90\%$ confidence level by advanced detectors assuming the spin axis of the recycled neutron star aligns with the total orbital angular momentum of the binary.
Second, using Bayesian inference, we show that it is possible to tell whether or not the spin axis of the recycled neutron star tends to be aligned with the binary orbit using $\gtrsim 30$ detections.
Finally, interesting constraints can be placed on neutron star magnetic field decay after $\gtrsim 300$ detections, if the spin periods and magnetic field strengths of
Galactic binary neutron stars are representative of the merging population.
\end{abstract}

\maketitle

\section{\label{sec:Intro}Introduction}
On August 17, the LIGO~\cite{aLIGO} and Virgo~\cite{aVirgo} detectors observed GW170817, the first gravitational-wave (GW) signal from a binary neutron star (BNS) merger~\cite{GW170817}.
The discoveries of an electromagnetic counterpart, visible across the electromagnetic spectrum, ushered in the highly-anticipated era of multimessenger astronomy~\cite{emfollow170817,GRB170817,170817optical,170817radio,emfollowOzGrav}.
Improved merger rate estimates now indicate that around 10--200 detections of BNS mergers are expected per year of operation at design sensitivities for Advanced LIGO and Virgo~\cite{LIGOrate10,GW170817,LVCprospects}.
This will enable detailed studies of the population properties of BNS systems.

In 1974, the first pulsar in a BNS system PSR B1913+16 was discovered in radio pulsar surveys \cite{HulseTaylor74,HulseTaylor75}.
There are now 18 BNS systems known in our Galaxy; see Ref. \cite{Tauris17} for a recent review and note that 4 new systems were discovered in the past year. Of particular significance is the Double Pulsar PSR J0737$-$3039A/B \cite{Burgay03,Lyne04}. It remains the only BNS system in which both stars can be observed as radio pulsars.
In the standard scenario, the recycled NS in a BNS system gets spun up by accreting matter and angular momentum from its companion~\cite{Tutukov73,Flannery75,Smarr76,Kalogera07,Postnov14LRR}.
The final spin period depends on the rate and duration of the accretion process.
There is a tendency for the spin periods of
the recycled pulsars to be shorter in
closer systems \cite{Tauris17}. This is because in a wider BNS progenitor system, the companion of the first born NS, usually a helium star, is more evolved when the mass transfer phase starts. Therefore, the first born pulsar generally has less time to gain angular momentum before its companion detonates as a supernova.

It is generally believed that NS spin down due to the loss of rotational energy by powering magnetically driven plasma winds~\cite{Goldreich69,Contop99,Spitkovsky06}.
Magnetic fields play a key role determining the astrophysical character of a NS (see~\cite{AshleyYuri17} and references therein).
Understanding the evolution of magnetic fields will give insights into both the physics of NS interiors and the observed population of pulsars in the Galaxy.
In this paper, we demonstrate that GW measurements of NS spins may shed new light on NS magnetic field evolution.

The spin of merging compact objects is imprinted on the gravitational waveform.
While there are six degrees of freedom describing the two spin vectors, the gravitational waveform depends primarily on a single combination of parameters called the effective spin parameter~\cite{IMR_Ajith11,Cutler93}
\begin{align}
\label{eq:chieff}
	\chi_\text{eff} \equiv
    \frac{m_{1}\chi_1\cos(\theta_1) + m_{2} \chi_2\cos(\theta_2)}{m_{1}+m_{2}} .
\end{align}
Here, $(\theta_1, \theta_2)$ are the angles between the spin vectors and the orbital angular momentum vector, $(m_1, m_2)$ are component masses, and $(\chi_1, \chi_2)$ are the dimensionless spin magnitudes of the merging objects~\cite{AbbottO1BNS,Dietrich15,Brown12NSaLIGO},
\begin{align}\label{eq:chi_aligned}
 \chi_{i} = c I_{i}\,\omega_{i}/Gm_{i}^2 .
\end{align}
Here, $i \in \{1,2\}$, $I$ is the NS moment of inertia, and $\omega$ is the angular spin frequency.
Physically, $\chi_\text{eff}$ is the mass-weighted sum of the spin projections along the axis of the orbital angular momentum. The presence of spins affects the gravitational-wave phase evolution.
As $\chi_\text{eff}$ is increased above zero, the merger takes place at a higher frequency, and produces a relatively longer signal in the observing band.
Conversely, as $\chi_\text{eff}$ is made increasingly negative, the merger takes place at a lower frequency, producing a relatively shorter signal.

In the case of stellar-mass binary black hole mergers, it has been shown that GW measurements of spin can be used to distinguish different formation channels (see, e.g., \cite{Vitale17,Stevenson17,Colm17,WillFarr17}). Black holes formed through isolated binary evolution are expected to spin in alignment with the binary orbital angular momentum, whereas the spin orientation of black holes in dynamically formed binaries is likely to be isotropically distributed. By combining multiple detections, it is possible to constrain the typical misalignment angle of black hole spins and infer the relative fraction of different subpopulations.

While published GW measurements of GW170817 are consistent with $\chi_\text{eff}=0$~\cite{GW170817}, advanced detectors operating at design sensitivity may be able to measure statistically significant spin in BNS inspirals.
In this paper, we point out that a significant fraction of future LIGO-Virgo detections may contain measurable signatures of spin.
Combining multiple measurements from an ensemble of BNS signals, it will be possible to constrain the {\em shape} of the $\chi_\text{eff}$ distribution from which we can infer interesting NS population properties.
In particular, we show that GW measurements can be used to (1) constrain the typical misalignment angle between the NS spin axis and the orbital angular momentum, (2) distinguish different NS equations of state (EOS) and (3) constrain the decay timescale of NS magnetic fields.

The organization of this paper is as follows.
In Sec.~\ref{sec:spinDistri}, we present fiducial models of $\chi_{\rm{eff}}$ distributions. We start from computing the effective spins of Galactic BNS systems at the time of binary mergers and outline relevant ingredients to build the models.
Additional details are provided in Appendices~\ref{sec:spin_BNSmerger} and~\ref{sec:BNSpopModel}. In Sec. \ref{sec:sen_chieff}, we study the sensitivity of advanced detectors to measure NS spins by performing parameter estimation analysis of realistic signal injections with the {\tt LALInference} package \cite{LALinference}. We also discuss the prospect of detecting $\chi_{\rm{eff}}$ in future BNS detections.
In Sec. \ref{sec:Bayes_results}, we present a Bayesian framework for inference of BNS population properties.
We demonstrate model selection and hyperparameter estimation using Monte-Carlo data.
Finally, in Sec. \ref{sec:conclu}, we provide concluding remarks.

\section{Fiducial models}
\label{sec:spinDistri}

\subsection{Expected effective spins for Galactic pulsar-neutron star systems at binary mergers}
\label{sec:BNSchi_radio}

There are 18 Galactic BNS systems observed in the radio band so far (for which details are given in Appendix \ref{sec:BNSpopModel}); 10 of these binaries are expected to merge within a Hubble time, ranging from about 46 Myr to 2.7 Gyr. We first look at what values of $\chi_{\rm{eff}}$ are expected for Galactic BNS at merger. Pulsar timing observations of these systems yield extremely precise measurements of pulsar spin periods, spin-down rates, orbital periods and eccentricities, and (in most cases) individual masses for both the pulsars and their companions. To compute at-merger $\chi_{\rm{eff}}$, three additional ingredients are needed.
\begin{table*}
\begingroup
\renewcommand{\arraystretch}{1.3}
 \begin{tabular}{lcccccccccccccc}
  \hline
  \hline
 Pulsar Name & $\chi_{\rm{eff}}^{{\rm{radio}}}$ & $\chi_{\rm{eff}}^{{\rm{low}}}$ & $\chi_{\rm{eff}}^{{\rm{decay}}}$ & $\chi_{\rm{eff}}^{{\rm{PAL1}}}$ & $\chi_{\rm{eff}}^{{\rm{high}}}$ & $\mathcal{R}$ & $M_c$ & $q$ & $m_p$ & $m_c$ & \multirow{2}{*}{$\Lambda_{p}$} & \multirow{2}{*}{$\Lambda_{c}$} & \multirow{2}{*}{$\tilde{\Lambda}$} & \multirow{2}{*}{Ref.} \\
 of BNS systems & ($\times 10^{-3}$) & ($\times 10^{-3}$) & ($\times 10^{-3}$) & ($\times 10^{-3}$) & ($\times 10^{-3}$) & (${\rm{Gyr}}^{-1}$) & ($M_{\odot}$) & & ($M_{\odot}$) & ($M_{\odot}$) & & & & \\
   \hline
  J1946+2052 & 15.3 & 14.3 & 14.5 & 20.0 & 20.3 & 22.0 & 1.0882 & 1.0 & 1.25 & 1.25 & 541.9 & 541.9 & 541.9 & \cite{BNS1946p20}\\
  J0737$-$3039A & 11.3 & 9.5 & 10.1 & 13.2 & 14.0 & 11.6 & 1.1253  & 0.933 & 1.3381 & 1.2489 & 358.3 & 544.7 & 442.8 & \cite{Cameron17BNS}\\
  J1757$-$1854 & 11.3 & 9.0 & 9.6 & 12.5 & 13.4 & 13.1 & 1.1893  & 0.960 & 1.3384 & 1.3946 & 357.8 & 276.5 & 314.8 & \cite{Kramer06Sci}\\
  J1913+1102 & 9.1 & 8.4 & 8.9 & 11.4 & 12.1 & 2.1 & 1.2390  & 0.756 & 1.64 & 1.24 & 92.3 & 568.4 & 243.1 & \cite{Ferdman17} \\
  B2127+11C & 8.1 & 4.5 & 6.0 & 6.2 & 8.3 & 4.6 & 1.1805  & 0.997 & 1.358 & 1.354 & 326.8 & 332.9 & 329.8 & \cite{Jacoby06}\\
  J1756$-$2251 & 9.1 & 4.2 & 8.3 & 5.8 & 11.6 & 0.6 & 1.1178  & 0.917 & 1.341 & 1.230 & 353.5 & 596.3 & 460.7 & \cite{Ferdman14}\\
  B1913+16 & 4.1 & 2.1 & 3.1 & 2.9 & 4.3 & 3.3 & 1.2309  & 0.964 & 1.4398 & 1.3886 & 225.3 & 284.1 & 253.3 & \cite{Weisberg10}\\
  B1534+12 & 6.5 & 1.9 & 5.6 & 2.6 & 7.9 & 0.4 & 1.1659  & 0.991 & 1.3330 & 1.3455 & 366.8 & 346.2 & 356.4 & \cite{Fonseca14} \\
  J0509+3801 & 3.1 & 1.4 & 2.5 & 2.0 & 3.5 & 1.7 & 1.2197 & 0.920 & 1.34 & 1.46 & 348.6 & 204.8 & 268.5 & \cite{Lynch0509}\\
  \hline
  \hline
  BNS Name & \multicolumn{2}{c}{$\chi_{\rm{eff}}$ ($\times 10^{-3}$)} & \multicolumn{2}{c}{$M_c$ ($M_{\odot}$)} & \multicolumn{2}{c}{$q$} & \multicolumn{2}{c}{$m_1$ ($M_{\odot}$)} & \multicolumn{2}{c}{$m_2$ ($M_{\odot}$)} & \multicolumn{3}{c}{$\tilde{\Lambda}$}\\
  \hline
  GW170817 & \multicolumn{2}{c}{($-$10, 20)} & \multicolumn{2}{c}{$1.188_{-0.002}^{+0.004}$} & \multicolumn{2}{c}{(0.7, 1.0)} & \multicolumn{2}{c}{(1.36, 1.60)} & \multicolumn{2}{c}{(1.17, 1.36)} & \multicolumn{3}{c}{$<=800$} & \cite{GW170817} \\
  \hline
  \hline
 \end{tabular}
 \endgroup
   \caption{Effective spins for nine Galactic BNS systems expected at merger under different assumptions of magnetic field decay and EOS (see text for details). Also included are the current spins derived from radio observations ($\chi_{\rm{eff}}^{{\rm{radio}}}$), the relative merger rate $\mathcal{R}=1/T_{c}$ with $T_c$ being the binary coalescence time; chirp mass $M_c$, mass ratio $q$ (defined to be $\leq 1$), masses for the pulsar $m_p$ and its companion $m_c$, the corresponding tidal deformability parameters $\Lambda_p$ and $\Lambda_c$ (assuming AP4 EOS) and $\tilde{\Lambda}$ which determines the leading contribution of tidal effects to gravitational-wave phase evolution. We assume $q=1$ for the recently discovered J1946+2052 for which only total mass is measured. For comparison, we also include relevant properties of GW170817. The range of each parameter is the 90\% confidence interval calculated assuming the low-spin prior.}
\label{tb:pulsarBNSchi}
\end{table*}

First, an EOS model is used to convert spin periods $P$ to spin magnitudes $\chi$. Our fiducial choice is the AP4 EOS but we also consider the PAL1 model (see \cite{Lattimer01EOS} and references therein).
We note that these two models both survive the $2 M_{\odot}$ pulsar constraint (see \cite{Lattimer12} for a recent review). For a NS mass of $1.4M_{\odot}$, the NS radius is 11 km and 14 km for AP4 and PAL1, respectively. The fastest spinning NS in a BNS system that will merge within a Hubble time, J1946+2052, has $P=17$ ms \cite{BNS1946p20}. Assuming a pulsar mass of $1.25 M_{\odot}$, this corresponds to a $\chi$ of 0.031 and 0.043 for AP4 and PAL1, respectively.

Second, we assume that NS spin down due to magnetic dipole braking\footnote{While it is incorrect that the spin-down is due to magnetic dipole radiation, it is the dipole that plays the key role and the dipole spin-down formula is correct to order unity.}. Our standard choice for NS magnetic field evolution is that the magnetic field remains constant over time. We also consider a model in which the magnetic field decays exponentially\footnote{This is used just as an example of the field evolution. Bransgrove et al. did not consider recycled pulsars in detail in their scenario (much colder crust with the field affected by accretion).} with a typical timescale $\tau_{B}=\unit[170]{Myr}$~\cite{AshleyYuri17}.
Magnetic field decay leads to higher $\chi_\text{eff}$ because the source of spin-down disappears. But the exact amount of increment depends on $\tau_{B}$ and merger time.
For example, PSR J1946+2052 has the shortest merger time $T_{c}=46$ Myr, the at-merger spin period is found to be 18.2 ms and 18.0 ms for the constant magnetic field and decaying magnetic field cases, respectively. On the other hand, for PSR B1534+12 the spin period would increase from its current 38 ms to only 44 ms if the magnetic field decays on a 170 Myr timescale; the final spin period would be 131 ms if its magnetic field remains constant over the course of 2.7 Gyr.

Third, we assume that the spin axis of the recycled NS is aligned with the total binary orbital angular momentum.
This is a plausible assumption given that the only measured misalignment angle (which comes from the Double Pulsar) is small: $<3.2^{\circ}$ at 95\% confidence, assuming that the observed emission comes from both
magnetic poles which is the scenario favored by radio data~\cite{FerdmanDPSR}. In an alternative to our fiducial models of $\chi_\text{eff}$ distribution that is described below, we also relax this assumption and adopt an isotropic distribution of spin orientations.

Throughout this paper, we consider the standard binary formation channel where a binary consists of a recycled NS and a nonrecycled NS. We further assume that nonrecycled NS possess negligible spins at the time of binary mergers. This is motivated by the following observations: 1) PSR J0737$-$3039B of the Double Pulsar has $P=2.8$ s which corresponds to $\chi \sim 10^{-4}$.
2) PSR J1906+0746 is another nonrecycled pulsar in a BNS system\footnote{It is noted that the possibility of a massive white dwarf companion can not be ruled out from radio observations. We take it as a BNS since its orbital characteristics and mass estimates are consistent with other Galactic BNS systems.}. Its current spin period (144 ms) corresponds to $\chi \approx 0.004$. Having a much higher spin-down rate ($\dot{P}=2\times 10^{-14}$) than PSR J0737$-$3039B ($\dot{P}=9\times 10^{-16}$), it is expected to merge with $\chi <10^{-3}$.

In Table \ref{tb:pulsarBNSchi} we list effective spins at merger for nine Galactic BNS that contain a recycled pulsar and that will merge within a Hubble time.
We calculate $\chi_\text{eff}$ four different ways:
\begin{itemize}
\item $\chi_{\rm{eff}}^{{\rm{low}}}$: no magnetic field decay and AP4 EOS
\item $\chi_{\rm{eff}}^{{\rm{decay}}}$: magnetic field decay with $\tau_{B}=\unit[170]{Myr}$ and AP4 EOS
\item $\chi_{\rm{eff}}^{{\rm{PAL1}}}$: no magnetic field decay and PAL1 EOS
\item $\chi_{\rm{eff}}^{{\rm{high}}}$: magnetic field decay with $\tau_{B}=\unit[170]{Myr}$ and PAL1 EOS
\end{itemize}
We also show the current spins $\chi_{\rm{eff}}^{{\rm{radio}}}$, which are calculated with AP4 EOS and thus can be directly compared with $\chi_{\rm{eff}}^{{\rm{low}}}$ and $\chi_{\rm{eff}}^{{\rm{decay}}}$.
Given the four different ways of calculating $\chi_\text{eff}$, we find that $\chi_{\rm{eff}}^{{\rm{low}}}$ yields the lowest spins, which range from $10^{-3}$ to 0.014.
The values of $\chi_\text{eff}^\text{PAL1}$ are $\approx40\%$ higher than those obtained for $\chi_\text{eff}^\text{low}$.
The distribution of $\chi_\text{eff}^\text{decay}$ depends on the interplay between two timescales of $T_{c}$ and $\tau_{B}$.
The optimistic case is $\chi_{\rm{eff}}^{{\rm{high}}}$: five out of nine systems are expected to have effective spins above 0.01 up to 0.02.
Note that the 90\% confidence interval of $\chi_{\rm{eff}}\in (-0.01,\,0.02)$ for GW170817 is completely consistent with pulsar observations of Galactic BNS systems.

\subsection{Fiducial models of BNS spin distribution}

In order to link GW measurements to NS evolution, we need to know $\pi(\chi_\text{eff}|M,\vec{\Xi})$: the  probability density function for the effective spin parameter $\chi_\text{eff}$ at merger given a model $M$ with population hyperparameters $\vec\Xi$.
The idea is that by measuring $\chi_\text{eff}$, we can make inferences about the NS population properties described by the hyperparameters $\vec\Xi$.
For example, we investigate our ability to constrain the NS magnetic field decay timescale $\tau_B\in\vec\Xi$. Here we present a suite of fiducial models that represent plausible distributions of $\pi(\chi_\text{eff}|M,\vec\Xi)$ for illustrative purposes. Such models result from a natural extension of different scenarios considered in the previous section for Galactic BNS.

\begin{figure}[h]
\centering
\includegraphics[width=\columnwidth]{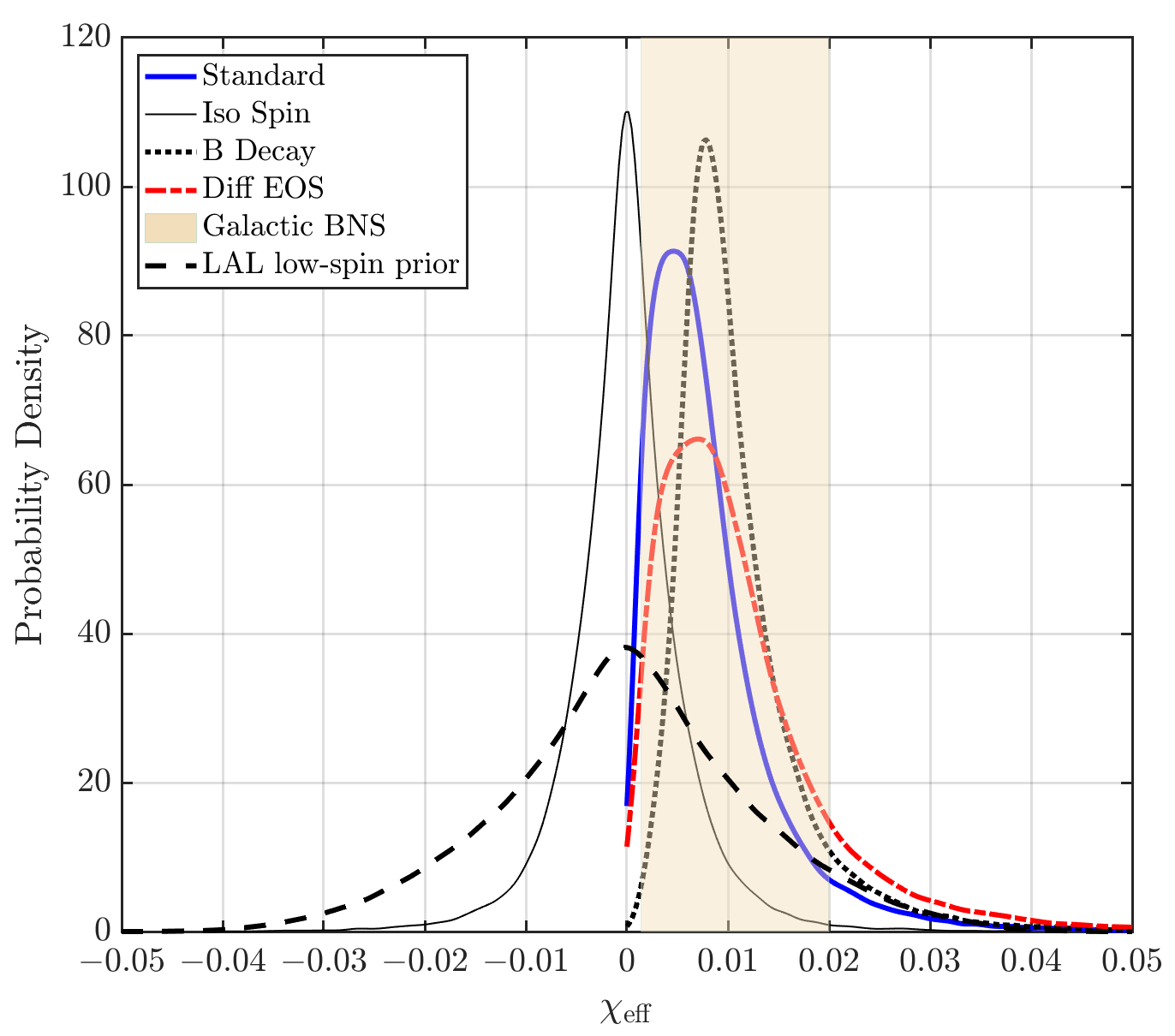}
\caption{
Probability density distribution of $\chi_{\rm{eff}}$ expected for a population of BNS inspiral events for four different models (Table \ref{tb:BNSmodel}), as compared to the low-spin prior usually used in gravitational-wave data analysis. The vertical shaded region indicates a plausible range expected for Galactic BNS systems extrapolated at merger (Table \ref{tb:pulsarBNSchi}).
}
\label{fig:Chieffsim1}
\end{figure}

To construct each fiducial model, we first construct the probability density function $\pi(\vec\xi_0)$ for a vector of parameters $\vec\xi_0$ describing the BNS system at the time of its birth.
These parameters include: masses for the recycled NS ($m_p$) and its companion ($m_c$), spin period $P$ and magnetic field strength ($B_0$) of the recycled NS, binary orbital period $P_b$ and eccentricity $e_0$.
This six-dimensional distribution is constructed using observations of known Galactic BNS systems (see Appendix ~\ref{sec:BNSpopModel} for details).
We generate Monte-Carlo samples using $\pi(\vec\xi_0)$ and follow the same procedure as described in Sec. \ref{sec:BNSchi_radio} to obtain the distribution of $\chi_{\rm{eff}}$ at merger.

We construct four fiducial models, labeled as {\sc Standard}, {\sc Iso Spin}, {\sc B Decay}, and {\sc Diff EOS}.
Each model corresponds to a different vector of hyperparameters $\vec\Xi$ as detailed in Table \ref{tb:BNSmodel}.

Our {\sc standard} model adopts the AP4 EOS and assumes a) the spin axis of the recycled NS is aligned with the total binary orbital angular momentum and b) the magnetic field decay time is long compared to the age of the Universe (i.e., no magnetic field decay).

The {\sc Iso Spin} model is identical to the {\sc Standard} model except we assume that $\cos\theta_1$ is drawn from a uniform distribution.
To be clear, we think that it is likely that $\cos\theta_1\approx1$ for the vast majority of BNS.
We put forward the {\sc Iso Spin} model as a means of falsifying this plausible assumption.

The {\sc B Decay} model is identical to {\sc Standard}, except that we assume an exponentially decaying magnetic field with a decay timescale $\tau_{B}=\unit[170]{Myr}$. The {\sc Diff EOS} model is identical to {\sc Standard}, except that we use the PAL1 EOS instead of AP4.

The distributions $\pi(\chi_\text{eff}|M)$ for our four fiducial models are shown in Fig.~\ref{fig:Chieffsim1}. For reference, the shaded vertical band indicates a plausible range of $\chi_{\rm{eff}}$ expected at merger for nine Galactic BNS systems listed in Table \ref{tb:pulsarBNSchi}.
The black dashed curve shows the low-spin prior used in the analysis of GW170817~\cite{GW170817}.
We use the same prior for our simulations presented in the next section.
We compute this prior by assuming a flat distribution in $(-0.05, \, 0.05)$ for $\chi_1$ and $\chi_2$, a uniform distribution for $\cos\theta_1$ and $\cos\theta_2$ between $-1$ and 1, and a uniform distribution for $q$ between 0.125 and 1. The shape of the $\chi_{\rm{eff}}$ prior is determined primarily by the first two assumptions with little dependency on the prior of $q$.

\begin{table}
 \begin{tabular}{lccc}
  \hline
 Model & Spin orientation & Magnetic field decay & EOS\\
   \hline
  {\sc Standard} & aligned & no & AP4 \\
  {\sc Iso Spin} & isotropic & no & AP4 \\
  {\sc B Decay} & aligned & yes, $\tau_{B}=170$ Myr & AP4\\
  {\sc Diff EOS} & aligned & no & PAL1 \\
  \hline
  \hline
 \end{tabular}
   \caption{Hyperparameter values of fiducial models.}
\label{tb:BNSmodel}
\end{table}
There are a number of interesting features worthy of remark in Fig.~\ref{fig:Chieffsim1}.
First, for all models except {\sc Iso Spin}, a significant fraction of BNS are characterized by relatively large spins with $\chi_\text{eff}>0.01$: 25\% for {\sc Standard} up to 45\% in {\sc Diff EOS}.
These results are consistent with Table \ref{tb:pulsarBNSchi}.

Second, we note that, for our four models, essentially all of the BNS possess $|\chi_\text{eff}|<0.05$. While this is consistent with the low-spin prior usually used in GW parameter estimation, it is worth noting that: 1) our models predict $\chi_\text{eff}>0$, except {\sc Iso Spin} which is used as an alternative. 2) Although quite unlikely, it is possible that recycled NS have $\chi>0.05$, with about 2\% probability in {\sc Standard} to 6\% probability in {\sc Diff EOS}. 3) Negligible spins are expected for nonrecycled NS.

Third, there are clear differences between different models. In the {\sc Standard} model, $\chi_{\rm{eff}}$ is found to be between 0 and 0.04 with $99.7\%$ probability. In the {\sc Iso Spin} model, the distribution is symmetric with respect to zero spin and $|\chi_{\rm{eff}}|\lesssim 0.02$. Both {\sc B Decay} and {\sc Diff EOS} predict generally higher values of $\chi_{\rm{eff}}$ than {\sc Standard}, extending the high end to 0.05. But they change the shape of $\pi(\chi_{\rm{eff}})$ in a different way. For {\sc Diff EOS}, the larger NS radii allowed by PAL1 slightly pull the distribution towards higher values. For {\sc B Decay}, the peak position gets shifted such that there is essentially no support for $\chi_{\rm{eff}}=0$. Although the differences among four models are subtle, they may be resolved with a sufficiently large number of events as demonstrated in Sec. \ref{model-selection}. In particular, the peak position of $\pi(\chi_{\rm{eff}})$ in {\sc B Decay} depends on $\tau_{B}$. We show in Sec. \ref{tau} that $\tau_{B}$ can be constrained after hundreds of detections.

\section{Measurement precisions of effective spins using advanced GW detectors}
\label{sec:sen_chieff}
In order to compare the distributions of $\chi_\text{eff}$ to the sensitivity of advanced detectors, we carry out Monte-Carlo simulations.
We assume the two-detector Advanced LIGO network operating at design sensitivity\footnote{We use the standard zero detuning high power sensitivity curve, which is publicly available at https://dcc.ligo.org/LIGO-T0900288/public. Due to the computational cost of {\tt LALInference} runs, we do not consider other advanced detectors. This only makes our conclusions more conservative.}.
Posterior distributions of signal parameters including $\chi_\text{eff}$ are obtained using the parameter estimation code {\tt LALInference} as part of the LSC Algorithm Library ({\tt \href{https://git.ligo.org/lscsoft/lalsuite/tree/lalinference_o2}{LALsuite}}). We use the Markov-Chain Monte-Carlo engine \cite{lalMCMC_Rover06,lalMCMC08} in {\tt LALInference} to perform stochastic sampling of signal parameter space.

We adopt the post-Newtonian TaylorF2 waveform model \cite{TaylorF2Vines11,TaylorF2Bohe13,TaylorF2Arun09,TaylorF2Miko05,TaylorF2Bohe15,TaylorF2Mishra16}, which allows us to simultaneously account for the effects of spins and tides.
Tidal effects are included via the dimensionless tidal deformability parameters $\Lambda_{1}$ and $\Lambda_2$. To the leading order, they affect the GW phase evolution via $\tilde{\Lambda}$, which is a mass-weighted combination of $\Lambda_{1}$ and $\Lambda_2$ and defined such that $\tilde{\Lambda}=\Lambda_{1}=\Lambda_{2}$ when $q=1$. In Table \ref{tb:pulsarBNSchi}, we list the values of tidal parameters for nine Galactic BNS systems assuming the AP4 EOS. Such calculations are carried out through the \texttt{lalsim-ns-params} functionality as part of the {\tt LALSimulation} package.

We perform the analysis in the frequency range 30--2048 Hz as in the case of GW170817 \cite{GW170817}. Our adopted waveform model includes only the inspiral part of the complete inspiral-merger-ringdown signal.
This is not expected to affect our results as the merger-ringdown occurs in a frequency band ($\gtrsim 1500$ Hz) where the detector sensitivity degrades significantly: the signal to noise ratio ($S/N$) decreases by $\lesssim 0.2\%$ when reducing the high frequency limit to 1024 Hz. For the same reason, the abrupt frequency cutoff in the waveform model as investigated in \cite{abrupt} does not have a significant impact on our results.

Our choices of priors are: 1) the low-spin prior depicted as the black curve in Fig. \ref{fig:Chieffsim1}; 2) uniform prior for component masses $0.6 M_{\odot}<m_{2}\leq m_{1}<5.8 M_{\odot}$ with further constraints such that $0.89M_{\odot}<M_{c}<1.63M_{\odot}$ and that $q>0.125$; 3) uniform priors for $\Lambda_{1,2}$ between 0 and 3000. Note that $\Lambda_{1,2}=0$ corresponds to black holes and the upper bound is beyond the 90\% confidence region found for GW170817 under the low-spin prior (see fig. 5 in \cite{GW170817}). 4) The prior for sky location assumes an isotropic distribution on the sky.

Our next aim is to obtain the approximate scaling law between realistic measurement uncertainties of $\chi_{\rm{eff}}$ and $S/N$.
We simulate a number of signals with component masses, spins (taking the conservative $\chi_{\rm{eff}}^{\rm{low}}$) and tidal parameters given in Table \ref{tb:pulsarBNSchi} for nine Galactic BNS. The luminosity distances are uniformly spaced between 100 and 300 Mpc.
We set other parameters to be identical for different signal injections.
The right ascension ($13^\text{h} \, 10^\text{m}$), declination ($-23^\circ \, 23'$), and GPS time (1187008882) are chosen to match those of GW170817.
The inclination angle $\iota$, the polarization angle $\psi$, and the phase at coalescence $\phi_c$ are all (somewhat arbitrarily) set to zero as they are not expected to impact on $\chi_{\rm{eff}}$ measurements.

\begin{figure}[h]
\centering
\includegraphics[width=\columnwidth]{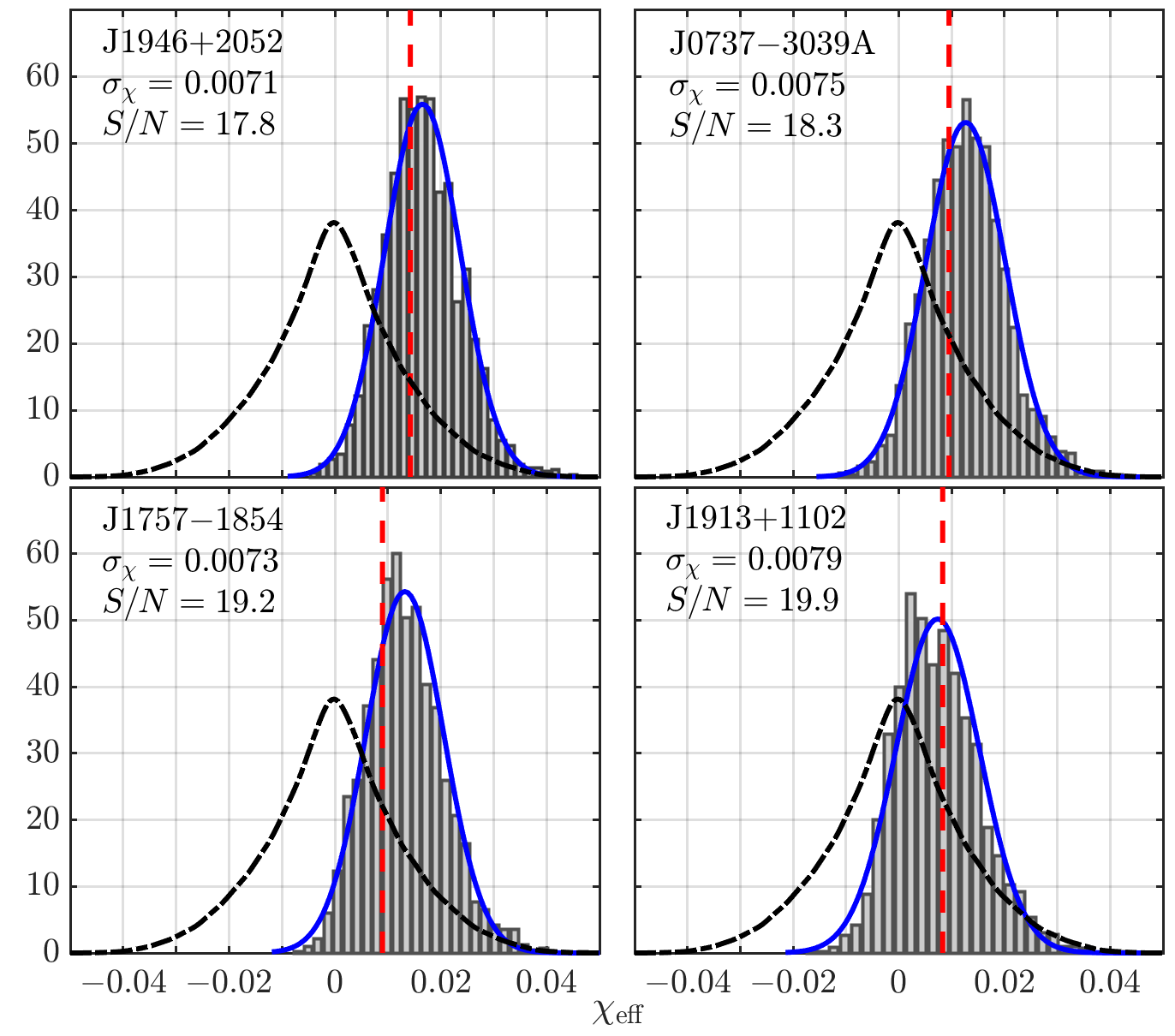}
\caption{Posterior probability density distribution of $\chi_\text{eff}$ generated by analyzing Monte-Carlo data using {\tt LALInference}.
The data consist of a BNS signal at $\unit[200]{Mpc}$ added to Hanford and Livingston design sensitivity noise. The blue curve is the Gaussian distribution with mean and standard deviation determined from posterior samples. The black dash-dotted curve is the prior distribution. The red vertical line marks the true injection value. In each panel we list the pulsar name of the Galactic BNS (see Table \ref{tb:pulsarBNSchi}), standard deviation ($\sigma_{\chi}$) of posterior samples and $S/N$ of each signal injection.}
\label{fig:posterior}
\end{figure}

As a first step, we inject signals into zero-noise data to obtain posterior distributions that are independent of specific noise realizations. Figure~\ref{fig:posterior} shows four example posterior distributions of $\chi_\text{eff}$ with injections at a distance of 200 Mpc. The signals correspond to the four Galactic BNS systems with the highest at-merger effective spins: PSR J1946+2052, PSR J0737$-$3039A, PSR J1757$-$1854, PSR J1913+1102.
In each panel of Fig.~\ref{fig:posterior}, the posteriors can be well approximated with a Gaussian distribution (blue curve) and peak very close to the true injection value (red vertical line). For a PSR J1946+2052 type of signal, $\chi_\text{eff}=0$ can be excluded with $\gtrsim 2\sigma$ confidence at 200 Mpc.
This clearly demonstrates that there are chances to detect NS spins in BNS inspirals. 
Spin is not quite detectable ($> 1.2\sigma$)  for both the Double Pulsar and PSR J1757$-$1854.
The effective spin of PSR J1913+1102  ($\chi_\text{eff}=0.0083$) is similar to measurement uncertainty ($\sigma_{\chi}=0.0079$).

\begin{figure*}[!tbp]
  \centering
{\includegraphics[width=\columnwidth]{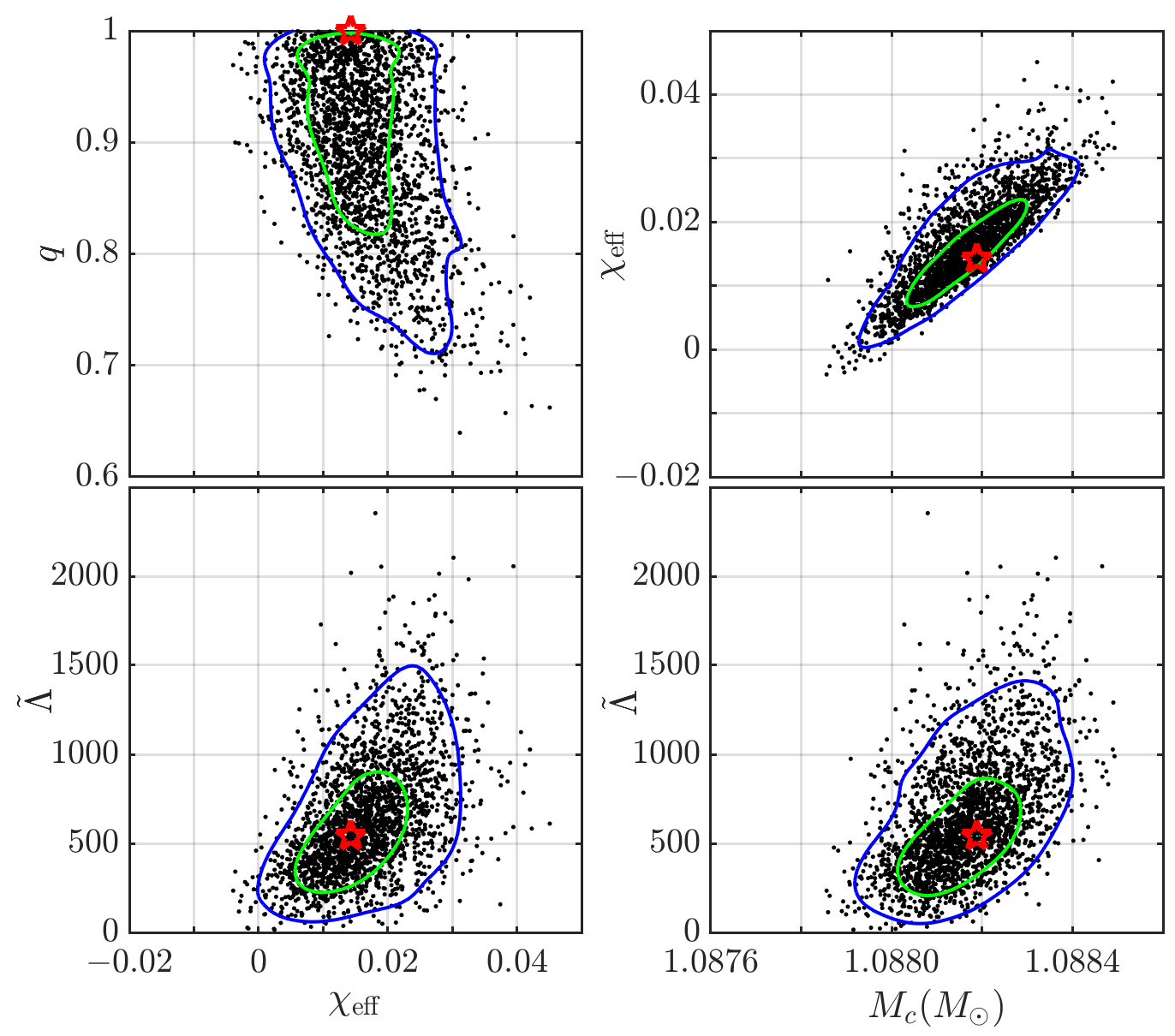}}
  \hfill
{\includegraphics[width=\columnwidth]{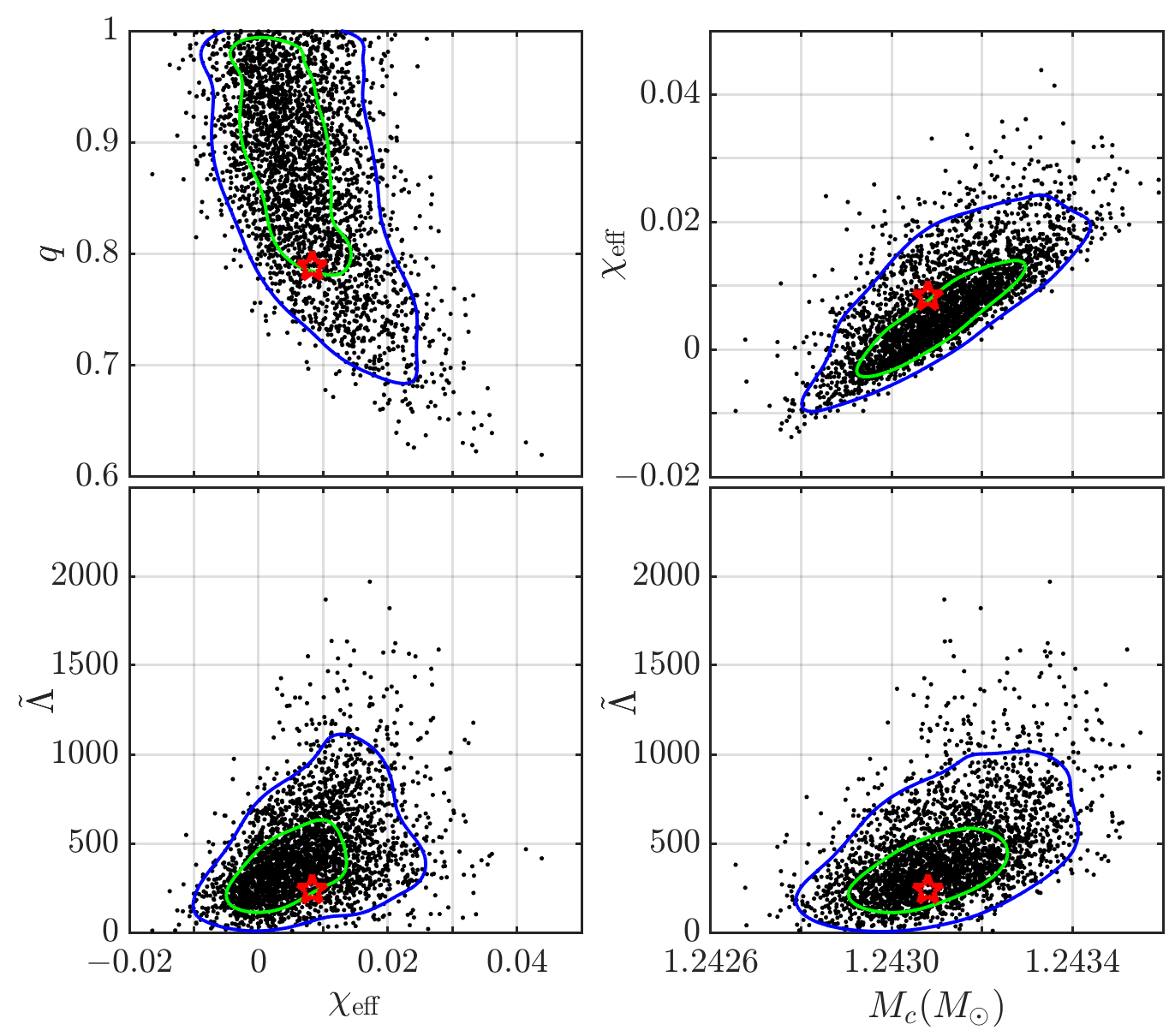}}
\caption{Marginalized 2-D posterior distributions for $\chi_\text{eff}$, $q$, $M_c$ and $\tilde{\Lambda}$ for two signal injections as in Fig. \ref{fig:posterior}: PSR J1946+2052 (left) and PSR J1913+1102 (right). The green and blue contours encompass 50\% and 90\% credible regions, respectively. The red stars mark true injection values and black dots are posterior samples returned from {\tt LALInferenceMCMC} analysis.}
\label{fig:post2d}
\end{figure*}
\begin{figure}[h]
\centering
\includegraphics[width=\columnwidth]{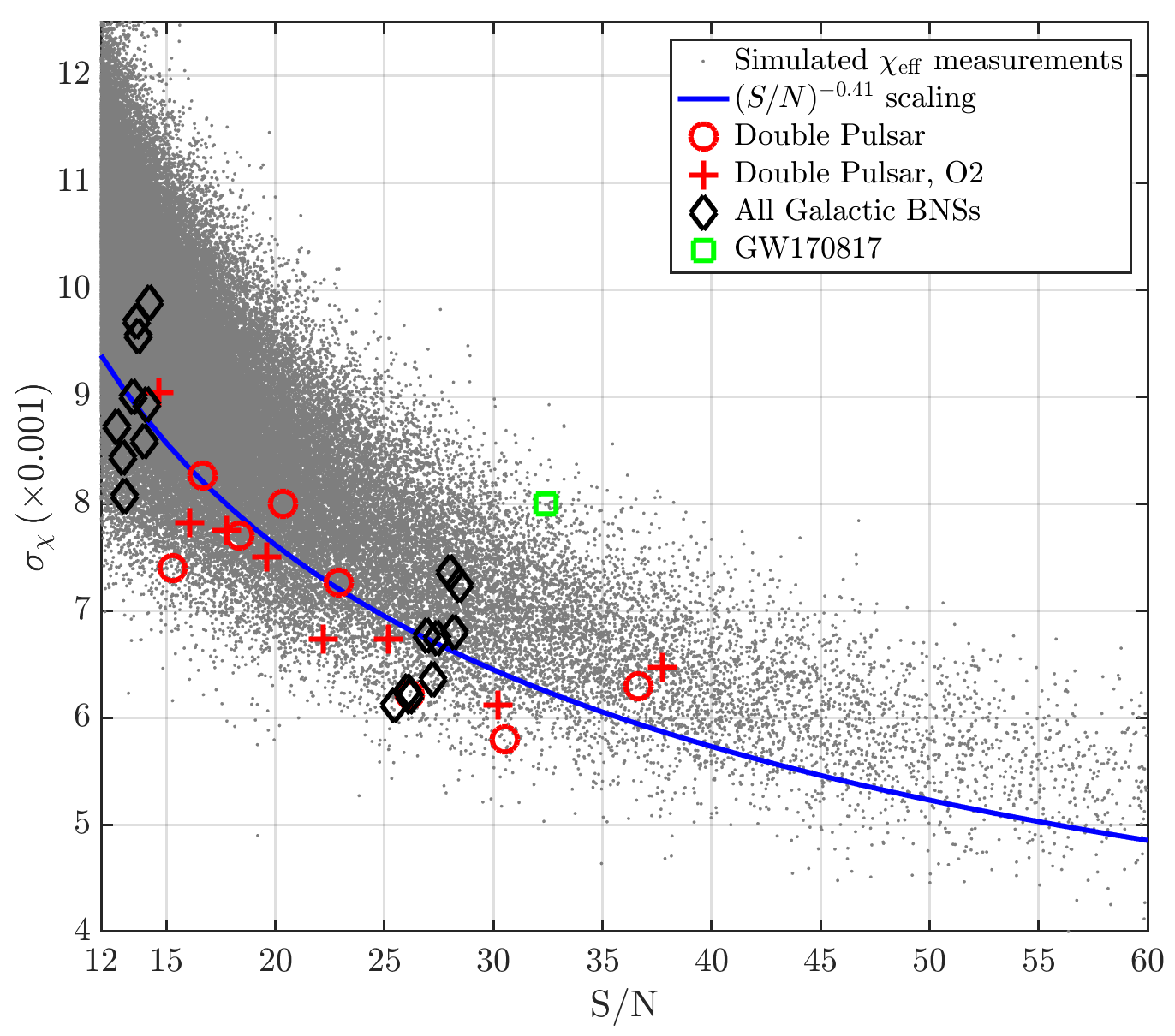}
\caption{The standard deviation ($\sigma_{\chi}$) of posterior distribution of $\chi_{\rm{eff}}$ as a function of signal to noise ratio ($S/N$). Shown here are measured by running parameter estimation code {\tt LALInference} on Monte-Carlo data at Advanced LIGO design (black diamonds and red circles) or O2 (red plus symbols) sensitivity. The blue curve is a fit for results found for injections to zero-noise data. Gray dots represent the distribution of simulated $\chi_{\rm{eff}}$ measurements used in Sec. \ref{sec:Bayes_results}. The green square marks the GW170817 uncertainty under the low-spin prior \cite{GW170817}.}
\label{fig:sigma}
\end{figure}

It is important to note that in Fig.~\ref{fig:posterior} the PSR J1913+1102 injection has the worst $\chi_\text{eff}$ measurement precision despite the highest $S/N$. 
This can be attributed to the correlations of $\chi_\text{eff}$ with other parameters.
In Fig. \ref{fig:post2d} we show the marginalized 2-D posterior distributions for $\chi_\text{eff}$, $q$, $M_c$ and $\tilde{\Lambda}$ for PSR J1946+2052 (left panel) and PSR J1913+1102 (right panel).
The two systems represent the two extreme ends in Table \ref{tb:pulsarBNSchi} for $q$, $M_c$ and $\tilde{\Lambda}$.
It can be seen from Fig. \ref{fig:post2d} that there are clear correlations for $\chi_\text{eff}$--$q$ and $\chi_\text{eff}$--$M_c$. The fact that PSR J1913+1102 has the lowest $q$, with the true value located in the long tail of $\chi_\text{eff}$--$q$ correlation, leads to the worst spin measurement among all nine Galactic BNS (placed at the same distance). This in turn results in a slightly worse chirp mass measurement due to the nearly linear $\chi_\text{eff}$--$M_c$ correlation than PSR J1946+2052. In contrast, we find $\chi_\text{eff}$--$\tilde{\Lambda}$ and $M_c$--$\tilde{\Lambda}$ are only weakly correlated.

In Fig.~\ref{fig:sigma}, we plot $\sigma_\chi$ measured from {\tt LALInference} simulations as a function of $S/N$.
We add signals to Gaussian noise characterized by Advanced LIGO design sensitivity.
For comparison with GW170817, we also include injections into noise consistent with LIGO's second observing run O2.
Red symbols are results for injections of the Double Pulsar type of signals placed at different distances. The measurement precisions with O2 sensitivity (red plus symbols) are found to be comparable to\footnote{This is valid when we set the low-frequency cutoff at 30 Hz. If we extend this cutoff to lower frequencies, some improvements may be achieved with design sensitivity.} that of design sensitivity (red circles) at the same $S/N$. We find realistic $\chi_{\rm{eff}}$ measurement uncertainties at Advanced LIGO design sensitivity can be described by the following relation
\begin{equation}
\sigma_{\chi}=\left(0.0086\pm 0.0008 \right)\left(\frac{S/N}{15}\right)^{-0.41}\, ,
\label{eq:sigma_SNR}
\end{equation}
where the exponent is from a fit to zero-noise simulations for the Double Pulsar, and the mean uncertainty and the scatter are found by extrapolating measurements of all Galactic BNS at design sensitivity (black diamonds in Fig.~\ref{fig:sigma}) to a $S/N$ of 15. We find systems with lower $q$ generally have larger measurement uncertainty for the same $S/N$ (see also Fig. \ref{fig:posterior}), whereas the scaling relations for different systems are very similar. The mean scaling relation is depicted by a blue curve in Fig.~\ref{fig:sigma}.

Interestingly, one obtains a different scaling relationship for binary black holes for which $\Lambda_i=0$. Repeating the same analysis using binary black hole waveforms \cite{IMRPhenomP}, we find $\sigma_{\chi}\propto (S/N)^{-1}$ given a fixed mass ratio ($q=0.933$). The slower improvement of $\chi_{\rm{eff}}$ measurement precisions while increasing $S/N$ for BNS can therefore be attributed to correlations between $\chi_{\rm{eff}}$ with tidal effects as shown in Fig. \ref{fig:post2d}. A detailed investigation of this feature will be included in a future work.

For comparison, we also show the actual measurement of GW170817 as a green square in Fig.~\ref{fig:sigma}. It lies above the blue curve by about 10\% after accounting for the scatter. This can be explained with the following factors: 1) calibration uncertainties of real data, which is typically $\lesssim 10\%$ \cite{GW170817}, and 2) subtraction of the glitch that was superposed on the signal in the LIGO-Livingston data could have impacted the uncertainties for parameter estimates.
In the next section, we scale up Eq. (\ref{eq:sigma_SNR}) by 10\% to be more compatible with GW170817. We show in Fig.~\ref{fig:sigma} gray dots as our simulated $\chi_{\rm{eff}}$ measurements. The probability distribution of $S/N$ is $p(S/N)\propto (S/N)^{-4}$ for events uniformly distributed in the detector's sensitive comoving volume with other parameters properly randomized \cite{Schutz11}. Note that events with a network matched-filter $S/N <$ 12 are excluded because they fall below the threshold necessary for unambiguous detection \cite{LVCsnr12}. The median $S/N$ in our samples is 15.

\subsection{Prospects of detecting NS spins}
\label{sec:fraction_spin}

\begin{figure}[h]
\centering
\includegraphics[width=\columnwidth]{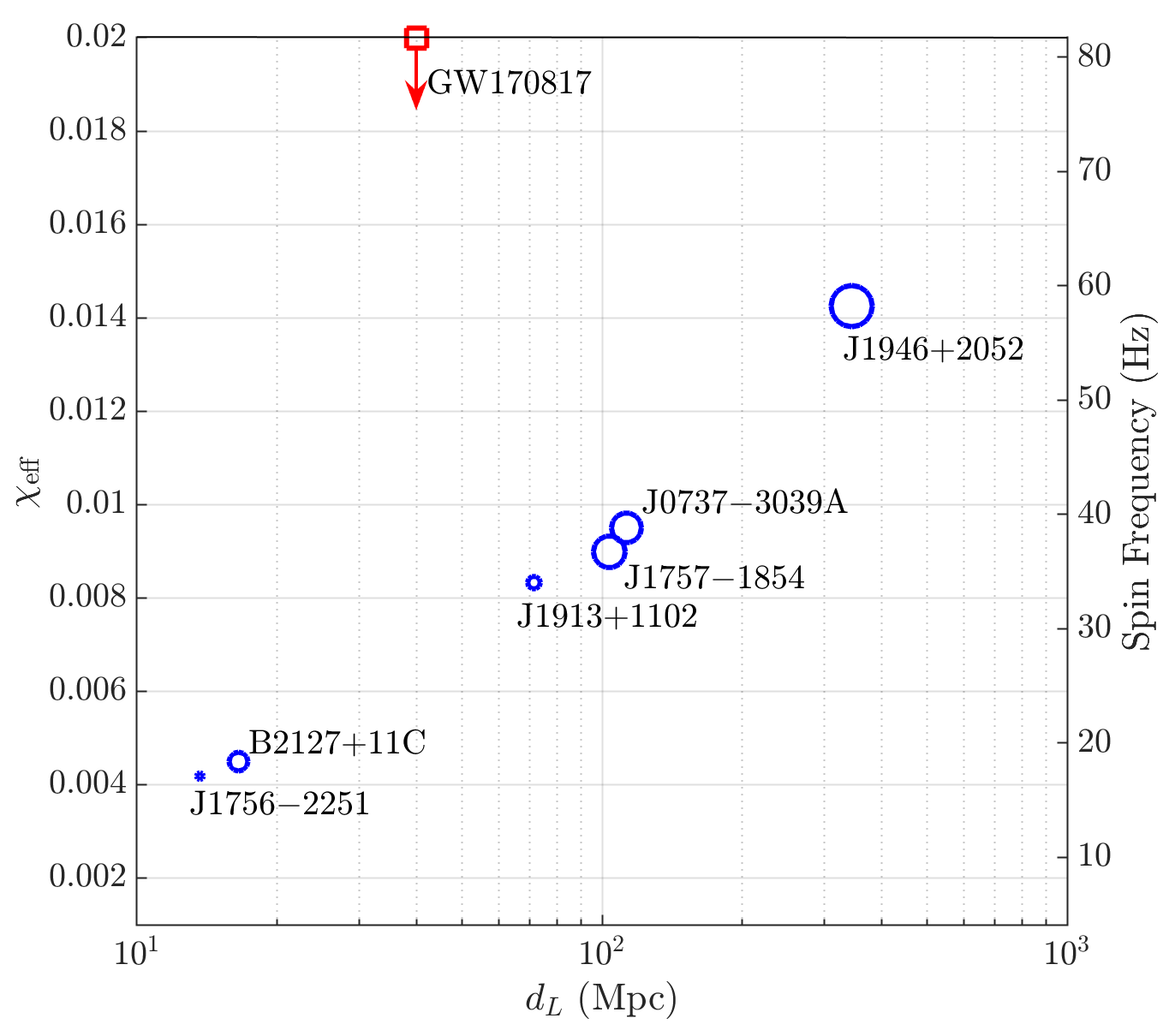}
\caption{A schematic diagram showing the sensitivity of Advanced LIGO to measure $\chi_\text{eff}$. $d_L$ is the approximate luminosity distance out to which $\chi_\text{eff}$ is measurable at the 90\% (1.6-$\sigma$) confidence level. Galactic BNS systems (blue circles) are placed at the sky location of GW170817 and assumed to be face-on (which allows the maximum $S/N$). Also shown is the published 90\% confidence upper limit for GW170817. A second y-axis shows the spin frequency of the recycled NS.}
\label{fig:BNSchi_dL}
\end{figure}

Here we investigate the prospect of detecting NS spins in GW events. First, we compare $\chi_\text{eff}$ measurement precisions derived previously with the effective spins of Galactic BNS systems at the time of merger (Table \ref{tb:pulsarBNSchi}).
In Fig. \ref{fig:BNSchi_dL}, we show the luminosity distance out to which $\chi_\text{eff}$ can be measurable at the 90\% (1.6-$\sigma$) confidence level for a two-detector Advanced LIGO network operating at design sensitivity.
Blue open circles are Galactic BNS systems, taking the conservative values of $\chi_\text{eff}^{\rm{low}}$ listed in Table \ref{tb:pulsarBNSchi}. We assume an optimal binary orientation ($\cos \iota =1$ which allows the maximum $S/N$) and all systems are placed at the sky location of GW170817.
To compute the distance, we extrapolate measurement uncertainties found in the previous section for injections at 200 Mpc onto zero-noise data using the $\sigma_{\chi}\propto (d_{L})^{0.41}$ relation to the distance that gives $\sigma_{\chi}=\chi_\text{eff}^{\rm{low}}/1.6$.
The size of the blue circles are proportional to the relative merger rate $\mathcal{R}$ (see Table \ref{tb:pulsarBNSchi}) of the particular system. The red square with downwards arrow indicates the 90\% confidence upper limit for GW170817.
We add a second y-axis for the spin frequency of the recycled NS in Fig. \ref{fig:BNSchi_dL} to relate $\chi_\text{eff}$ to radio observations.
The conversion from $\chi_\text{eff}$ to the spin frequency employs the average mass of six recycled pulsars shown in the plot (1.37 $M_{\odot}$) and $\chi_\text{eff}=\chi_{1}/2$, i.e., assuming $\chi_{2}=0$ and $m_{1}=m_{2}$; see Eqs. (\ref{eq:chieff}-\ref{eq:chi_aligned}).

To properly compute the fraction of future BNS detections with measurable effective spins, we simulate BNS events, each characterized by three quantities: $\chi_\text{eff}$, $S/N$ and $\sigma_{\chi}$.
We generate Monte-Carlo samples of $\chi_\text{eff}$ according to distribution models illustrated in Fig. \ref{fig:Chieffsim1}, and of $S/N$ according to $p(S/N)\propto (S/N)^{-4}$. We assign each event with a Gaussian measurement uncertainty given by Eq. (\ref{eq:sigma_SNR}) but scaled up by $10\%$. Gray dots in Fig.~\ref{fig:sigma} represent $10^4$ simulated events.
We find the fraction of events with $\chi_\text{eff}\geq 1.6 \sigma_{\chi}$ to be 13\% for {\sc Standard}, 20\% for {\sc B Decay}, and 25\%  for {\sc Diff EOS} with typical fluctuations of about 2\%.
This is in good agreement with Fig. \ref{fig:BNSchi_dL}. For PSR J1946+2052 types of systems, spins are measurable just beyond 300 Mpc. For spins represented by the Double Pulsar and PSR J1757$-$1854, we can probe up to about 100 Mpc. A naive estimate is one out of nine systems should possess measurable $\chi_\text{eff}$ in the \textit{standard} case. However, our fiducial models treat each Galactic system as equally representative of the true BNS population.
Given the relatively higher merger rate for these three fastest-spinning systems (as indicated by the size of the blue circles in Fig. \ref{fig:BNSchi_dL}), it is possible that a larger fraction of merging BNS have measurable spins.

\section{Bayesian inference}
\label{sec:Bayes_results}

In this section, we present a Bayesian formalism to infer BNS population properties from GW measurements of $\chi_\text{eff}$.
In Sec.~\ref{model-selection}, we demonstrate how GW measurements from an ensemble of detections can be used to carry out model selection, thereby allowing us to differentiate between the different models in Fig. \ref{fig:Chieffsim1}.
In Sec.~\ref{tau}, we show how to constrain the magnetic field decay timescale using a large number of detections.

\subsection{Model selection}\label{model-selection}
Let ${\cal L}(h|\vec\vartheta)$ be the likelihood of GW data $h$ given BNS parameters $\vec\vartheta$.
The GW parameter space includes 17 or more parameters including, e.g., inclination angle, luminosity distance, and component masses.
Here, however, we are only concerned with $\chi_\text{eff}$.
Thus, we marginalize over every parameter except $\chi_\text{eff}$ to obtain a marginalized likelihood function ${\cal L}(h|\chi_\text{eff})$.

The Bayesian evidence for a model $M$ given the data $h$ is
\begin{align}\label{eq:evidence}
{\cal Z}(h|M) =
\int d\chi_\text{eff} \,
 {\cal L}(h|\chi_\text{eff}) \,
\pi(\chi_\text{eff}|M) ,
\end{align}
where $\pi(\chi_\text{eff}|M)$ is the conditional prior distribution for $\chi_\text{eff}$ given a model $M \in$ \{{\sc Standard}, {\sc Iso Spin}, {\sc B Decay}, {\sc Diff EOS}\}. Rewriting Eq.~\ref{eq:evidence} in terms of posterior samples~\cite{MacKay}, we obtain
\begin{align}\label{eq:pos-samp}
{\cal Z}(h|M) =
\sum_{i=1}^{n_k}
\frac{\pi(\chi_\text{eff}^i|M)}{\pi(\chi_\text{eff}^i|\text{LAL})} .
\end{align}
Here, the index $i$ runs over $n_k$ posterior samples, each corresponding to a different value of $\chi_\text{eff}$; see also Eq.~(7) in~\cite{Colm17}.
The distribution $\pi(\chi_\text{eff}|\text{LAL})$ is the prior used by {\tt LALInference} in order to obtain the initial set of posterior samples.
It appears in Eq.~\ref{eq:pos-samp} in order to convert the posterior distribution output by {\tt LALInference} into a likelihood.

The total evidence from $N$ detections is simply the product of the evidence for each detection:
\begin{align}
{\cal Z}_\text{tot}(\{h\}|M) = \prod_{k=1}^N {\cal Z}(h_k|M) .
\end{align}
The Bayes factor (BF) comparing model $M_1$ and model $M_2$ is given by
\begin{align}
\text{BF}_{1,2} = \frac{{\cal Z}_\text{tot}(h|M_1)}{{\cal Z}_\text{tot}(h|M_2)} .
\end{align}
Following convention \cite{MacKay}, we impose a threshold of $\ln(\text{BF})=8$ ($\sim 3.6 \sigma$) to define the point beyond which one model is significantly favored over another.

In order to demonstrate model selection, we carry out Monte-Carlo simulations as described in Sec. \ref{sec:fraction_spin}.
To speed up the analysis, instead of running each signal injection with {\tt LALInference} which is computationally very expensive, we assume the likelihood function ${\cal L}(h|\chi_\text{eff})$ is described by a Gaussian distribution with width $\sigma_\chi$
\begin{align}\label{eq:L_approximate}
{\cal L}(h|\chi_\text{eff}) \propto N(\chi_\text{eff}^\text{ML}, \sigma_\chi) \, ,
\end{align}
where the mean $\chi_\text{eff}^\text{ML}$ is the maximum-likelihood estimate for a particular random noise realization.
Each value of $\chi_\text{eff}^\text{ML}$ is generated by adding a normally distributed random number (variance = $\sigma_\chi^2$) to the true value of $\chi_\text{eff}$.
We use the scaling relation between $\sigma_\chi$ and $S/N$ described in Sec. \ref{sec:sen_chieff} and visually depicted as gray dots in Fig.~\ref{fig:sigma}.

\begin{figure}[h]
\includegraphics[width=\columnwidth]{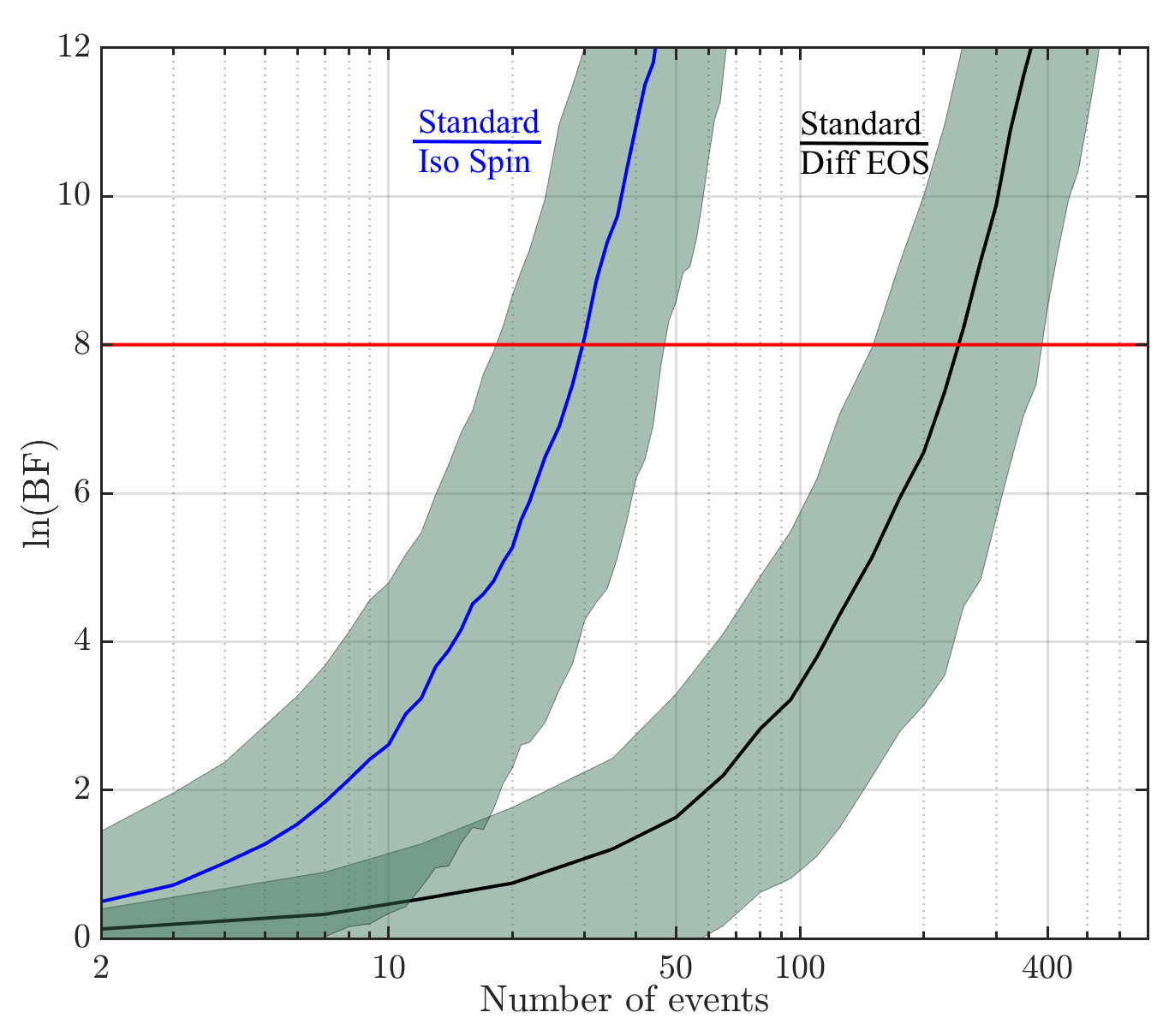}
\caption{Bayes Factor (BF) as a function of the number of events between models {\sc Standard} (the true underlying model), {\sc Iso Spin}, {\sc Diff EOS}. The shaded areas correspond to 1-$\sigma$ confidence regions.}
\label{fig:BFmodelSelect}
\end{figure}

Then, we calculate two Bayes factors (plotted in Fig.~\ref{fig:BFmodelSelect} as a function of the number of events): {\sc Standard / Iso Spin} (blue) and {\sc Standard / Diff EOS} (black).
The colored lines show the median value while the shaded area indicates the 1-$\sigma$ confidence region.

When $\ln\text{BF}_{1,2}\gtrsim 8$, we are able to clearly distinguish between the {\sc Standard} model and the two alternatives considered here.
The blue curve shows that advanced detectors can distinguish between {\sc Standard} and {\sc Iso Spin} after $\approx 30$ events. The black curve shows that advanced detectors can distinguish between {\sc Standard} and {\sc Diff EOS} after $\approx 250$ events. The Bayes factor of {\sc Standard / B Decay} is similar to that of {\sc Standard / Diff EOS} shown here. We study this model in a different way below.

\subsection{Constraining magnetic field decay}\label{tau}
Given an EOS, one can construct a posterior for the magnetic field decay timescale $\tau_B$
\begin{align}
\label{eq:posterior_tau}
p(\tau_{B}|h,M) \propto\,  & \pi(\tau_{B})
\int d\chi_\text{eff} \,
{\cal L}(h|\chi_\text{eff}) \pi(\chi_\text{eff}|M,\tau_{B})\, .
\end{align}
Here $\pi(\tau_{B})$ is the prior distribution for $\tau_{B}$, which is assumed to be log-uniform on the interval $(\unit[1]{Myr},\,\unit[10^4]{Myr})$. Note that such a wide prior is used only as an example to test our method and does not fold in any theoretical or observational information. Our primary purpose here is to demonstrate how many GW detections are required to place interesting constraints on $\tau_B$ using GW data alone.
The distribution $\pi(\chi_\text{eff}|M,\tau_B)$ is the conditional prior for $\chi_\text{eff}$ given model $M$ and the decay timescale $\tau_B$.
An example conditional prior is represented visually in Fig.~\ref{fig:Chieffsim1} for model {\sc B Decay} with $\tau_B=170$ Myr.

Marginalizing over $\chi_\text{eff}$, we obtain a posterior on $\tau_B$.
In practice, we do not carry out an integral over a continuous variable; we sum up posterior samples.
The discrete version for Eq.~(\ref{eq:posterior_tau}) is
\begin{align}
p(\tau_{B}|h,M) \propto \pi(\tau_{B})
\sum_{i=1}^{n} \frac{\pi(\chi_\text{eff}^i|M,\tau_{B})}{\pi(\chi_\text{eff}^i|\text{LAL})} .
\end{align}
Generalizing to $N$ events, the posterior for $\tau_B$ becomes
\begin{align} \label{eq:poster_tauB}
p(\tau_{B}|h,M) \propto \pi(\tau_{B})
\prod_{k=1}^N
\sum_{i=1}^{n_k} \frac{\pi(\chi_\text{eff}^{i,k}|M,\tau_{B})}{\pi(\chi_\text{eff}^{i,k}|\text{LAL})} .
\end{align}
Here, $\chi_\text{eff}^{i,k}$ represents the $i^\text{th}$ posterior sample from the $k^\text{th}$ event.

Following the same procedure described in Sec.~\ref{model-selection}, we generate an ensemble of detections. Events are generated using the {\sc B Decay} model with the true value of $\tau_{B}=170$ Myr. In order to infer $\tau_B$ using Eq. (\ref{eq:poster_tauB}), we generate $\pi(\chi_\text{eff}|M,\tau_B)$ for a vector of $\tau_B$ equally spaced in log space between 1 Myr and 10 Gyr.

\begin{figure}[h]
\includegraphics[width=\columnwidth]{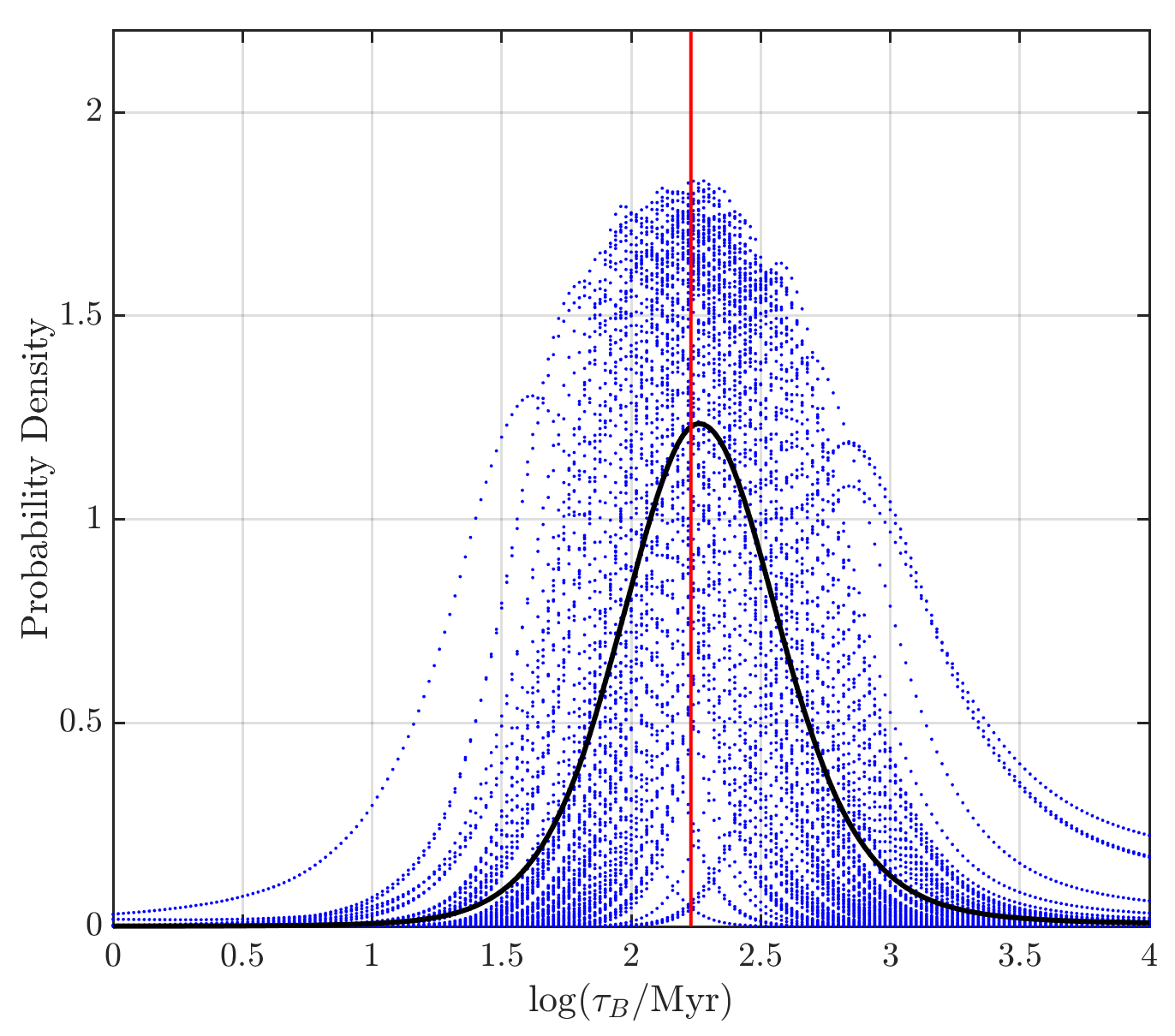}
\caption{The mean posterior distribution (black solid curve) and individual distributions for 100 random noise realizations (blue dotted curves) of the magnetic field decay timescale $\tau_{B}$ recovered using 300 events. The red vertical line marks the true injection value.}
\label{fig:pdftauB}
\end{figure}

Figure \ref{fig:pdftauB} shows the posterior distribution $p(\tau_{B}|h,M)$ inferred from 300 events.
The true value $\tau_B=\unit[170]{Myr}$ is marked with a red vertical line.
We repeat the simulation 100 times to illustrate cosmic variance.
In each realization, we draw a new set of values for $\chi_\text{eff}$ and $S/N$ and assign new measurement uncertainties from gray dots from Fig.~\ref{fig:sigma}; the corresponding posterior distributions are shown as blue dotted curves. The black solid curve shows the mean posterior from 100 simulations.
For $\approx 95\%$ of noise realizations, the true value of $\tau_B$ is recovered within the 2-$\sigma$ confidence interval. In this example, we find that $\tau_B$ can be constrained to be between $\unit[30]{Myr}$ and $\unit[1]{Gyr}$ with $\gtrsim 95\%$ confidence after 300 detections.

\section{Discussion and Conclusions}
\label{sec:conclu}
The detection of the BNS inspiral event GW170817 has opened up new opportunities to study NS physics.
In the next five years, observations of similar events will become increasingly more frequent as Advanced LIGO/Virgo approach their design sensitivities and new detectors such as KAGRA \cite{KAGRA} and LIGO-India \cite{LIGO-India} start to operate.
As advanced detectors observe tens, and eventually hundreds, of BNS inspirals, it will be possible to make inferences about NS population properties.
We demonstrate that GW measurements of spin will have implications for the typical spin tilt angle of NS after tens of detections and for the NS magnetic field evolution or EOS after hundreds of observations.

In this proof-of-concept analysis, we rely on a set of fiducial models.
The models are constructed to be consistent with radio pulsar observations, but they do not necessarily represent the full range of allowable parameter space.
While the fiducial models are suitable for illustrative purposes, the next step in this program is to develop a more nuanced model, which {\em does} allow for the full range of possible initial conditions. It should be possible to then combine all available radio and GW data within a single Bayesian framework, yielding optimal constraints on NS evolution.

We find that at least one and up to three out of nine known Galactic BNS systems that will merge within a Hubble time have spins detectable at 90\% confidence by Advanced LIGO operating at design sensitivity.
The true fraction of BNS with measurable spins may be higher since short-lived binaries like these three systems are less likely to be observed in radio.
Our fiducial model does not account for selection effects associated with the detection of radio pulsars in BNS systems. We assume that Galactic BNS detected in radio are representative of merging binaries detectable with GWs.
In future studies, it will be necessary to carefully include all selection effects in a formulation that attempts to combine GW and radio data.

In this work, we focus on BNS systems formed via the isolated binary evolution channel.
It is generally believed that there is insufficient time for the recycled NS to reach ms spin periods and the second NS spins down much faster and thus makes no contribution to effective spins at merger. While our fiducial models predict that nearly all merging BNS have $|\chi_{\rm{eff}}|<0.05$, binaries formed via dynamical captures could have much larger spins. Measurements of NS spins outside our fiducial model predictions will have interesting implications about their formation history.

\begin{acknowledgements}
We would like to thank the anonymous referees for helpful comments. We also thank Ryan S. Lynch, Matthew Bailes, Salvatore Vitale, Colm Talbot, Simon Stevenson and Christopher Berry, members of the LSC CBC working group and Rory Smith, Alexander Heger, Bernhard M\"uller, Sylvia Biscoveanu, Xi-Long Fan, Lijing Shao, Shuxu Yi and Gang Wang for useful conversations.
X.Z., E.T. \& Y.L. are supported by ARC CE170100004.
E.T. is supported through ARC FT150100281. S.O. acknowledges support through the ARC Laureate Fellowship Grant No. FL150100148.
P.D.L. is supported through ARC FT160100112 and ARC DP180103155.
This is LIGO Document LIGO-P1700400.
\end{acknowledgements}

\begin{appendix}
\section{\label{sec:spin_BNSmerger}NS spin evolution and BNS inspirals}
Here we describe the details to calculate final spin periods of NS from an initial set of BNS parameters as outlined in Sec. \ref{sec:spinDistri}.

The coalescence time of a circular binary due to the emission of GWs is \cite{Thorne87}
\begin{equation}
\label{eq:Tmerge}
T_{\rm{c,circ}}=\frac{5c^5}{256(2\pi f_{0})^{8/3}(G M_{c})^{5/3}},
\end{equation}
where $f_{0}$ is the initial binary orbital frequency, $M_c$ is the chirp mass defined as $M_c = M \eta ^{3/5}$, with $M= m_1+m_2$ being the total mass, and $\eta = m_1 m_2 /M^{2}$ is the symmetric mass ratio. The coalescence time for an eccentric binary can be computed as \cite{Peters-Mathews63,Peters64}
\begin{equation}
\label{eq:TmergeEcc}
T_{\rm{c}}=\frac{15c^5}{304}(2\pi f_{0})^{-8/3}(G M_{c})^{-5/3}a_{0}^{4}\int_{0}^{e_{0}}g(e){\rm{d}}e.
\end{equation}
Here $e_0$ is the initial binary orbital eccentricity, the constant $a_0$ and the function $g(e)$ are
\begin{equation}
\label{eq:a0}
a_{0}=(1-e_{0}^2)e_{0}^{-12/19}\left(1+\frac{121}{304}e_{0}^2\right)^{-870/2299},
\end{equation}
\begin{equation}
\label{eq:gofe}
g(e)=e^{29/19}\left(1-e^2\right)^{-3/2}\left(1+\frac{121}{304}e^2\right)^{1181/2299}.
\end{equation}
For an eccentric binary, the orbital frequency and eccentricity coevolve as
\begin{equation}
\label{eq:f0fe}
\frac{f(e)}{f(e_0)}=\left[\frac{\sigma(e_{0})}{\sigma(e)}\right]^{3/2},
\end{equation}
where the function $\sigma(e)$ is defined as
\begin{equation}
\label{eq:sigma2}
\sigma(e)=\frac{e^{12/19}}{1-e^2}\left(1+\frac{121}{304}e^2\right)^{870/2299}.
\end{equation}
Evolving the BNS systems listed in Table \ref{tab:BNSspin} that will merge within a Hubble time forward in time, we find that they are expected to be in nearly circular orbits ($e \lesssim 10^{-5}$) when they enter the sensitive frequency band ($f> 10$ Hz) of ground-based interferometers in between 46 Myr and 2.7 Gyr.

Assuming magnetic dipole braking, the NS spin frequency evolution follows \cite{Ostriker69}:
\begin{equation}
\label{eq:omegadot}
\dot{\omega}=-K\omega^{3},
\end{equation}
where $\omega=2\pi/P$ with $P$ being spin period, $\dot{\omega}={\rm{d}}\omega/{\rm{d}}t$. The torque parameter $K$ is defined as \cite{Spitkovsky06}
\begin{equation}
\label{eq:Kbraking}
K=\frac{B^{2}R^{6}}{c^{3}I}\left(1+\sin^{2}\alpha\right),
\end{equation}
where $B$ is the surface magnetic field strength, $R$ is the NS radius which is determined by the EOS for a given mass, $I$ is the moment of inertial, $\alpha$ is the misalignment angle between the magnetic dipole moment and spin axis. There are a few notes worthy of mentioning here. First, Eq. (\ref{eq:Kbraking}) is a convenient approximation to numerical results found in \cite{Spitkovsky06}. Second, there is a degeneracy between $B$ and $\alpha$. Since pulsar measurements collected here only allow us to infer $K$, we choose to fix $\alpha$ at $30^{\circ}$. The exact value of $\alpha$ is not important as the overall spin-down behavior is determined by $K$. Third, the spin-down estimator of $B$ can be derived from $P\dot{P}=4\pi^{2}K$, where $P$ and $\dot{P}$ are the spin period and period derivative, respectively.

The moment of inertial $I$ can be computed as
\begin{equation}
\label{eq:inertial}
I\simeq 0.237 mR^{2}\left[1+4.2\frac{m\, {\rm{km}}}{M_{\odot}R}+90\left(\frac{m\, {\rm{km}}}{M_{\odot}R}\right)^{4}\right].
\end{equation}
The above equation is found as an empirical relation by fitting to a sample of EOS \cite{Lattimer05}.

Assuming $K$ remains a constant throughout the binary's lifetime, it is straightforward to show that the spin angular frequency at coalescence is:
\begin{equation}
\label{eq:omegaf}
\omega_{f}=\frac{\omega_{i}}{\sqrt{1+2T_{\rm{c}}|\dot{\omega}_{i}|/\omega_{i}}},
\end{equation}
where $\omega_{i}$ and $\dot{\omega}_{i}$ are the initial values of angular spin frequency and its time derivative, respectively.

It is possible that the NS magnetic filed decays exponentially over time \cite{Romani90,Geppert94,Konar97,Cumming04,AshleyYuri17}
\begin{equation}
\label{eq:Bdecay}
B(t)=B_{0}e^{-t/\tau_B},
\end{equation}
where $\tau_B$ is the magnetic field decay timescale. Under this decaying-B field model, the final spin frequency is
\begin{equation}
\label{eq:omegafB}
\omega_{f}=\frac{\omega_{i}}{\sqrt{1+\tau_{B} \left[\exp(-2T_{\rm{c}}/\tau_{B})-1\right]\dot{\omega}_{i}/\omega_{i}}}.
\end{equation}

\begin{table*}
 \begin{tabular}{lcccccccccccc}
  \hline
  Pulsar Name & $P$ (ms) & $P_{\rm{f}}$ (ms) & $\chi_{i}$ & $\chi_{\rm{f}}$ & $P_b$ (days) & $e_0$ & $T_{\rm{c}}$ (Gyr) & Age (Gyr) & B ($10^9$ G) & $m_p$ ($M_{\odot}$) & $m_c$ ($M_{\odot}$) & Ref.\\
  \hline
  \multicolumn{13}{| c |}{systems will merge within a Hubble time}\\
  \hline
  J1946+2052 & 16.96 & 18.23 & 0.031 & 0.028 & 0.078 & 0.064 & 0.046 & 0.292 & 0.7 & \multicolumn{2}{c}{$M_{\text{tot}}=2.50 M_{\odot}$}  & \cite{BNS1946p20}\\
  J1757$-$1854 & 21.50 & 27.09 & 0.023 & 0.018 & 0.184 & 0.606 & 0.076 & 0.129 & 1.3 & 1.338 & 1.395 & \cite{Cameron17BNS} \\
 J0737$-$3039A  & 22.70 & 27.17 & 0.022 & 0.018 & 0.102 & 0.088 & 0.086 & 0.204 & 1.1 & 1.338 & 1.249 & \cite{Kramer06Sci}\\
 J0737$-$3039B & 2773 & 4596 & 0 & 0 &  0.102 & 0.088 & 0.086 & 0.049 & 260 & 1.249  & 1.338 & \cite{Kramer06Sci}\\
 J1913+1102 & 27.29 & 29.59 & 0.016 & 0.015 & 0.206 & 0.090 & 0.475 & 2.688 & 0.4 &  1.64 & 1.24 & \cite{Lazarus16} \\
  J1756$-$2251  & 28.46 & 61.95 & 0.017 & 0.008 & 0.320 & 0.181 & 1.656 & 0.443 & 0.9 & 1.341 & 1.230 & \cite{Ferdman14}\\
  B2127+11C  & 30.53 & 54.90 & 0.016 & 0.009 & 0.335 & 0.681 & 0.217 & 0.097 & 2.2 & 1.358 & 1.354 & \cite{Jacoby06}\\
  B1534+12  & 37.90 & 131.5 & 0.013 & 0.004 & 0.421 & 0.274 & 2.734 & 0.248 & 1.7 & 1.333 & 1.346 & \cite{Fonseca14}\\
 B1913+16  & 59.03 & 114.7 & 0.008 & 0.004 & 0.323 & 0.617 & 0.301 & 0.108 & 4.1 & 1.440 & 1.389 & \cite{Weisberg10}\\
 J0509+3801 & 76.54 & 166.86 & 0.006 & 0.003 & 0.380 & 0.586 & 0.574 & 0.153 & 4.3 & 1.34 & 1.46 & \cite{Lynch0509}\\
 J1906+0746  & 144.1 & 7536 & 0.003 & 0 & 0.166 & 0.085 & 0.308 & 0.0001 & 289 & 1.291 & 1.322 & \cite{vanLeeuwen15}\\
  \hline
   \multicolumn{13}{| c |}{systems will not merge within a Hubble time}\\
     \hline
 J1807$-$2500B  & 4.19 & 150.7 & 0.117 & 0.003 & 9.957 & 0.747 & 1044 & 0.81 & 0.1 & 1.366 & 1.206 & \cite{Lynch12}\\
 J1518+4904  & 40.93 & 788.7 & 0.012 & 0.001 & 8.634 & 0.249 & 8832 & 23.85 & 0.2 & \multicolumn{2}{c}{$M_{\text{tot}}=2.72 M_{\odot}$} & \cite{Janssen08}\\
   J1829+2456  & 41.01 & 95.9 & 0.013 & 0.005 & 1.176 & 0.139 & 55.35 & 12.38 & 0.2 & \multicolumn{2}{c}{$M_{\text{tot}}=2.59 M_{\odot}$}   & \cite{Champion05}\\
 J0453+1559  & 45.78 & 885.1 & 0.010 & 0.001 & 4.072  & 0.113 & 1453 & 3.90 & 0.6 & 1.559 & 1.174 &  \cite{Martinez15}\\
 J1411+2551 & 62.45 & 423.6 & 0.008 & 0.001 & 2.616 & 0.170 & 465.9 & 10.35 & 0.4 & \multicolumn{2}{c}{$M_{\text{tot}}=2.538 M_{\odot}$} & \cite{BNS1411}\\
  J1753$-$2240 & 95.14 & - & 0.005 & 0 & 13.64 & 0.304 & $3\times 10^{4}$ & 1.55 & 1.7 & - & - & \cite{Keith09} \\
 J1811$-$1736  & 104.2 & 3262 & 0.004 & 0 & 18.78 & 0.828 & 1794 & 1.83 & 1.6 & \multicolumn{2}{c}{$M_{\text{tot}}=2.57 M_{\odot}$}  & \cite{Corongiu07}\\
 J1930$-$1852 & 185.5 & - & 0.003 & 0 & 45.06 & 0.399 & $5\times 10^{5}$ & 0.16 & 9.8 & \multicolumn{2}{c}{$M_{\text{tot}}=2.59 M_{\odot}$}  & \cite{Swiggum15}\\
  \hline
 \end{tabular}
 \caption{Spin periods and dimensionless spin magnitudes at the present ($P$ and $\chi_{i}$) and upon coalescence ($P_{\rm{f}}$ and $\chi_{\rm{f}}$) for pulsars in known Galactic BNS systems. $\chi<10^{-3}$ is listed as 0 and $P_{\rm{f}}>10$ s is considered as irrelevant in this table. We assume that the spin evolution is determined by magnetic dipole braking, the magnetic field does not decay and use the AP4 EOS. Also included are binary orbital period ($P_b$), orbital eccentricity ($e_0$), coalescence time $T_c$, characteristic age ($P/2\dot{P}$ where $\dot{P}$ is the observed spin period derivative), magnetic field strength ($B$), the mass of the pulsar ($m_p$) and its companion ($m_c$). The total mass ($M_{\text{tot}}$) is listed when no definitive component mass measurement is available; in this case equal-mass is assumed for calculations. For PSR J1753$-$2240 we assume 1.34 and 1.23 $M_{\odot}$ for $m_p$ and $m_c$, respectively. Note that PSR J1906+0746 is a nonrecycled NS and its companion's mass is included in the fit for mass distribution of the recycled NS.}
  \label{tab:BNSspin}
\end{table*}

\begin{table}
 \begin{tabular}{lcccc}
  \hline
  Parameter & Min & Max & Distribution &  Note\\
   \hline
  $P\, (\text{ms})$  & 4 & 62 & Normal & \multirow{2}{*}{Eq. (\ref{eq:P0Pd})}\\
  $P_{b}\, (\text{day})$ & 0.01 & 0.5 & Uniform & \multirow{2}{*}{Eqs. (\ref{eq:f0fe}-\ref{eq:sigma2}) }\\
  $e_0$  & 0.004 & 0.78 & Uniform & \\
  $\log(B_{0}/\text{G})$ & - & - & $N(9.0, 0.4)$ & AP4 \\
  $\log(B_{0}/\text{G})$ & - & - & $N(8.8, 0.4)$ & PAL1 \\
  \multirow{2}{*}{$m_{1}\, (M_{\odot})$} & \multirow{4}{*}{1.0} & \multirow{4}{*}{2.0} & $0.63\times N(1.34, 0.04)$ & \multirow{4}{*}{-}\\
  & & & $+0.37\times N(1.48, 0.13)$ & \\
  $m_{2}\, (M_{\odot})$ & & & $0.52 \times N(1.23, 0.06)$ & \\
    & & & $+0.48\times N(1.37, 0.05)$ & \\
  \hline
  \hline
 \end{tabular}
   \caption{Summary of the birth distributions used to construct our fiducial BNS models described in Sec. \ref{sec:spinDistri}. $P$ and $B_0$ are the initial spin period and magnetic field strength for the recycled NS, respectively. $P_b$ and $e_0$ are the birth binary orbital period and eccentricity, respectively. $m_1$ and $m_2$ are mass for the recycled NS and the secondly born NS, respectively. The two-Gaussian models for $m_1$ and $m_2$ distributions are obtained by fitting to mass measurements listed in Table \ref{tab:BNSspin}. $P$ follows a Normal distribution along the line of linear $\log P_{b} - \log P$ correlation given in Eq. (\ref{eq:P0Pd}).}
\label{tb:BNSpop}
\end{table}

\section{\label{sec:BNSpopModel}Construction of initial distributions}

In this section we describe the construction of our models for the distribution of BNS parameters at birth $\pi(\vec\xi_0)$. The models are fully motivated by pulsar observations of currently known Galactic BNS systems. These observations are summarized in Table \ref{tab:BNSspin}, including current and final (i.e., when binary merges) spin periods, the corresponding dimensionless spin magnitudes, binary orbital period and eccentricity, coalescence time, characteristic age and surface magnetic field strength, masses of the pulsar and its companion.

In our fiducial model, each binary consists of a recycled NS and a ``normal'' NS. Here ``normal" means that it is born with a period of $\gtrsim\unit[10]{ms}$ but quickly spins down to $\mathcal{O}(1)$ seconds period over $\sim\unit[10]{Myr}$~\cite{pulsarHandbook}.
PSR J1906+0746 is considered to be normal because it is a young ($\sim 100$ kyr) pulsar with much higher spin-down rate $2\times 10^{-14}$ \cite{Lorimer1906}.
Table \ref{tb:BNSpop} summarizes initial distributions. We describe the details below.

\begin{figure}[h]
\includegraphics[width=\columnwidth]{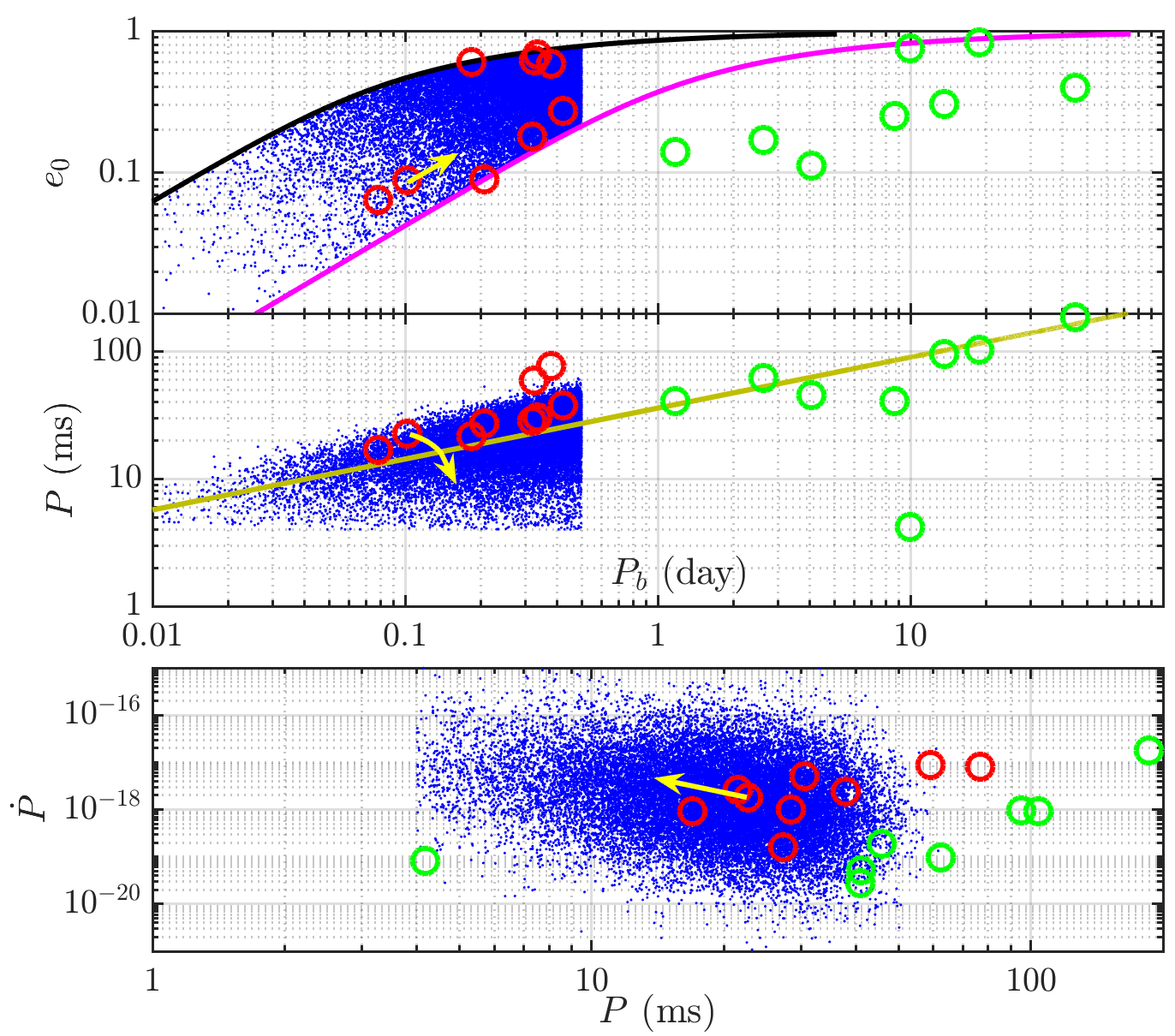}
\caption{Simulated BNS at birth (blue dots), along with known Galactic binary pulsar systems (open circles; only recycled pulsars are shown). Red (green) circles are for binaries that will (will not) merge within a Hubble time. For illustration, yellow arrows indicate plausible evolutionary tracks from the current state towards the initial state at the birth of the second NS for the Double Pulsar.}
\label{fig:PBP0e0sim}
\end{figure}

We first derive a model for NS mass distribution from 11 BNS systems that have precise component mass measurements. To represent the full range of mass ratio between the recycled NS and its companion, we fit separately a two-Gaussian model to 11 measurements. For the recycled NS, the main peak of the mass distribution is centered around 1.34 $M_{\odot}$ and the second around 1.48 $M_{\odot}$, with weight of approximately $60\%$ and $40\%$, respectively. For the nonrecycled NS, two comparable Gaussian components are centered around 1.23 $M_{\odot}$ and 1.37 $M_{\odot}$, respectively. Given these mass distributions, $90\%$ of our simulated BNS have $0.8\leq q \leq 1$. Note that we treat the recycled NS instead of the more massive one as the primary star, so it is possible that the primary is less massive, which corresponds to $\approx 25\%$ of all BNS. This fraction is in excellent agreement with Table \ref{tab:BNSspin} where 3 out of 11 binaries with definitive component mass measurements have a less massive recycled NS.

We assume $P_b$ is uniformly distributed in the range of $(0.01,\,0.5)$ days. This roughly corresponds to the range of measured orbital periods of Galactic BNS systems that will merge within a Hubble time, except that we extend to below the shortest orbital period of these binaries. Current searches for radio pulsars in relativistic binaries are limited to orbital periods longer than a few hours. The exact value of the lower cut for $P_b$ has no significant impact as the population is dominated by systems in wider orbits. We assume that the initial orbital eccentricity ($e_0$) is also uniformly distributed.

Figure \ref{fig:PBP0e0sim} provides an intuitive picture about the initial distributions of BNS parameters in our fiducial model. Blue dots mark the positions of synthetic BNS systems, whereas open circles represent known Galactic systems. Red (green) circles are for binaries that will (will not) merge within a Hubble time.

In the top panel of Fig. \ref{fig:PBP0e0sim}, we show the the distribution in the $P_{b}-e_{0}$ space for our simulated BNS systems. In the GW-driven regime, $P_{b}$ and $e_{0}$ coevolve following equations (\ref{eq:f0fe}-\ref{eq:sigma2}). This is indicated by magenta and black curves, which cross J1913+1102 and J1757$-$1854, respectively. We use these two curves to exclude binaries with either very short birth orbital periods and high $e_0$ (top-left corner) or long periods and low $e_0$ (bottom-right corner), as both are shown to be unlikely based on pulsar observations. Magenta and black curves, together with $P_b$ limits, lead to a range of $e_0$ between 0.004 and 0.8.

In the middle panel of Fig. \ref{fig:PBP0e0sim}, open stars indicate positions of recycled pulsars in BNS systems in the $P_{b}-P$ diagram. As one can see, there exists a linear correlation for $\log P_{b} - \log P$
\begin{equation}
\label{eq:P0Pd}
P=36 \pm 14 \, {\rm{ms}}\, (P_{b}/{\rm{days}})^{0.4}.
\end{equation}
The above equation is found by an empirical fit to 11 Galactic BNS systems in~\cite{Tauris17} while placing particular weight on the widest binary J1930$-$1852 and including qualitative differences of supernova progenitors for different BNS. Such a linear correlation is further confirmed by 16 Galactic BNS systems shown in this plot (excluding PSR J1807$-$2500B). The mean correlation is depicted as a straight line in the middle panel of Fig. \ref{fig:PBP0e0sim}. It lies below all red circles as it describes the relation at the birth of the second NS when slightly longer orbital periods and shorter spin periods are expected. We apply a lower cut-off of 4 ms for $P$ as informed by PSR J1807$-$2500B.

% In order to obtain the initial values of $P_b$, $P$, $\dot{P}$ and $e_0$ for 7 Galactic BNS systems at the birth of the second NS, we evolve these systems back in time assuming magnetic dipole braking and GW-driven orbital decay. The time is set as half of the characteristic age. Although this is somewhat arbitrary, it should be noted that the spin period approaches zero when evolving the pulsar back by the full amount of its characteristic age. The evolution is shown as yellow tracks in the bottom panel of Fig. \ref{fig:PBP0e0sim} and Fig. \ref{fig:PPdotsim}. The upper bound (black line) in the $P_{b}$-$P_{1}$ plane is set by PSR J1756$-$2251 with initial orbital period of 0.37 day and spin period of 43 ms. The lower bound (red line) is set by PSR J1757$-$1854 with initial orbital period of 0.32 day and spin period of 15 ms. The black and red lines, together with the cut-off in orbital period, determine the maximum and minimum values for $P_1$ to be 49 and 4 ms respectively.

The magnetic field strength of the recycled NS at the time of the birth of its companion is assumed to follow a log-Normal distribution with a mean value $\mu_{\rm{log},B}=9.0\, (8.8)$ and a standard deviation $\sigma_{\rm{log},B}=0.4$ for the AP4 (PAL1) EOS. Such a distribution is found to give a reasonable $P-\dot{P}$ diagram for recycled pulsars while taking into account the fact that pulsars generally have shorter spin periods and higher spin-down rates at an earlier time. This is demonstrated in the bottom panel of Fig. \ref{fig:PBP0e0sim}. Because PAL1 allows relatively large NS radii, the required magnetic field strength to produce the same spin-down rate is lower; see Eq. (\ref{eq:Kbraking}). This illustrates the covariance between the EOS and the magnetic field in probing NS spin evolution. Tighter constraints on the EOS will enable better understanding on NS magnetic field evolution through spin measurements.

% For the second-born NS, we assume that i) the initial spin period follows a Gaussian distribution with a mean of $\unit[300]{ms}$ and a standard deviation of $\unit[150]{ms}$ \cite{FKaspi06,RidleyLorimer10}; and ii) the initial magnetic field strength follows a log-Normal distribution with a mean $\mu_{\rm{log},B}=12.65$ and a standard deviation $\sigma_{\rm{log},B}=0.55$ \cite{FKaspi06}. Note that our results are insensitive to details of these two distributions as in our fiducial model the second NS makes no contribution to $\chi_{\rm{eff}}$ unless the magnetic field decay timescale is much shorter than binary coalescence time.

% Finally, we briefly comment on directions to improve the fiducial models presented here.
% Our primary goal in developing these fiducial models was to cover parameter ranges informed by pulsar observations.
% Some prescriptions are therefore ad-hoc.
% We assume the magnetic field strength remains a constant when evolving the recycled pulsars back in time to obtain the initial distributions of spin periods and period derivatives.
% However, we allow magnetic field to decay when evolving $\pi(\vec\lambda_0)$ {\em forward} in time to obtain at-merger distribution $\pi(\vec\lambda_m)$.
% While this inconsistency does not significantly impact our results, future models should include a self-consistent treatment of magnetic field decay.

\end{appendix}

\bibliography{ref}

\end{document}